\documentstyle[12pt,epsf,a4]{article}
\renewcommand{\thefootnote}{\fnsymbol{footnote}}
\begin{document}    

\title{\vskip-3cm{\baselineskip14pt
\centerline{\normalsize\hfill DESY 00--124}
\centerline{\normalsize\hfill TTP00--14}
\centerline{\normalsize\hfill hep-ph/0009029}
\centerline{\normalsize\hfill July 2000}
}
\vskip.7cm
{\tt MATAD}: a program package for the computation of MAssive TADpoles
\vskip.3cm
}
\author{
{Matthias Steinhauser}
  \\[3em]
  { II. Institut f\"ur Theoretische Physik,}\\ 
  { Universit\"at Hamburg, D-22761 Hamburg, Germany}
  \\[.5em]
  \centerline{and}
  \\[.5em]
  { Institut f\"ur Theoretische Teilchenphysik,}\\
  { Universit\"at Karlsruhe, D-76128 Karlsruhe, Germany}
}
\date{}
\maketitle

\begin{abstract}
\noindent
In the recent years there has been an enormous development in the evaluation
of higher order quantum corrections.
An essential ingredient in the practical calculations is provided by vacuum
diagrams, i.e. integrals without external momenta.
In this paper a program package is described which can deal with one-, two-
and three-loop vacuum integrals with one non-zero mass parameter.
The principle structure is introduced and the main parts of the package are
described in detail. Explicit examples demonstrate the fields of application.
\end{abstract}

\thispagestyle{empty}
\newpage
\setcounter{page}{1}

\renewcommand{\thefootnote}{\arabic{footnote}}
\setcounter{footnote}{0}


\noindent
{\bf PROGRAM SUMMARY}

\begin{itemize}

\item[]{\it Title of program:}
  {\tt MATAD}

\item[]{\it Available from:}\\
  {\tt
  http://www-ttp.physik.uni-karlsruhe.de/Progdata/MATAD/1/
  }

\item[]{\it Computer for which the program is designed and others on which it
    is operable:}
  Any work-station or PC where {\tt FORM} is running.

\item[]{\it Operating system or monitor under which the program has been
    tested:} 
  UNIX, {\tt FORM}~2.3

\item[]{\it No. of bytes in distributed program including test data etc.:}
  $706000$

\item[]{\it Distribution format:} 
  ASCII

\item[]{\it Keywords:} 
  three-loop computations, vacuum integrals, computer algebra,
  automation of computations

\item[]{\it Nature of physical problem:}
  Multi-loop integrals are needed for the evaluation of 
  quantum corrections. An important class of loop diagrams 
  is covered by so-called vacuum integrals which have no external
  momentum. {\tt MATAD} can analytically compute those 
  one-, two- and three-loop vacuum integrals where one mass scale 
  is present.

\item[]{\it Method of solution:}
  The method of integration-by-parts is used in order to obtain 
  recurrence relations which reduce complicated integrals to a small
  set of so-called master integrals. They have to be evaluated once
  and for all.
  In addition a user interface is provided which makes it easy to put in
  complicated diagrams in a rather compact way.
  
\item[]{\it Restrictions on the complexity of the problem:}
  The restrictions on the complexity are given by the hardware limitations
  of the computer and the limits on the size of the
  storage files inside {\tt FORM}.

\item[]{\it Typical running time:}
  The runtime strongly depends on the complexity of the diagram under
  consideration. It may vary form a few seconds to the order of a few
  weeks.

\end{itemize}


\newpage


\noindent
{\bf LONG WRITE-UP}

\section{Introduction}

The high experimental precision reached at the electron-positron
machines LEP (CERN) and SLC (SLAC) and the
hadron collider TEVATRON (FERMILAB)
requires from the theoretical side the evaluation of higher
order quantum corrections.
In the cases where perturbative methods are applied the quantum corrections
can be expressed through an expansion in the coupling constant of the
underlying theory. The individual terms can in turn be expressed
through so-called Feynman diagrams, which are often classified as multi-leg
or multi-loop diagrams. Vacuum integrals, i.e. integrals without external
momenta, constitute an important sub-class and often serve as
building blocks in complex calculations.

In general the momentum integration of the loop integrals is divergent in
four space-time dimensions. At present the most practical method to cope with 
this problem in higher loop orders is
based on Dimensional Regularization~\cite{dimreg}.
There, the four space-time 
dimensions are replaced by $D=4-2\varepsilon$ dimensions. Then
the integrals are solved for a choice of $\varepsilon$ that renders 
them finite. Finally an expansion for $\varepsilon\to0$ is performed
and the divergences manifest themselves as poles in $\varepsilon$.

Important progress in practical computations
has been made roughly 20 years ago by establishing an algorithm for
the evaluation of 
propagator-type diagrams up to three loops in the massless
case~\cite{CheTka81}.
They are important if there is only one external momentum which
sets the mass scale for the problem.
The formulae have been implemented
on the computer in the {\tt FORM}~\cite{form}
package {\tt MINCER}~\cite{mincer}.
In 1995 for the first time three-loop diagrams in the opposite limit,
i.e. zero external momentum but massive lines, were 
systematically examined~\cite{Bro92}.
Usually these are denoted as vacuum or tadpole diagrams.
In~\cite{Bro92} all integrals contributing to the photon propagator
have been considered. The main characteristics of this class of diagrams is
that the massive line forms a closed loop.
These considerations have been extended to the $W$ boson
current correlators which led to one of the most prominent applications
of three-loop vacuum integrals, namely the $\rho$ parameter at ${\cal
  O}(\alpha\alpha_s^2)$~\cite{Avdrho,CheKueSte95rho}.
The remaining cases have been considered
in~\cite{Avd97,Bro98,FleKal99,CheSte00}.
Thus it is --- at least in principle --- possible to treat all 
problems where exactly one heavy mass is involved.

\begin{figure}[t]
  \begin{center}
    \epsfxsize=14cm
    \leavevmode
    \epsffile[75 580 500 740]{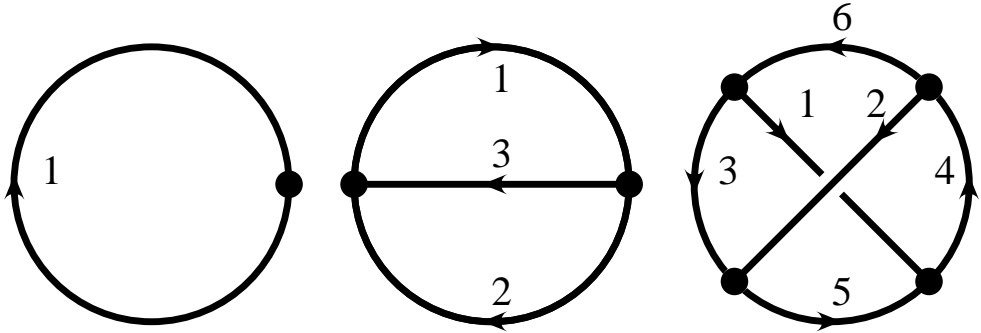}
    \caption{\label{fig:tad123}Prototype topologies for 
      one-, two- and three-loop vacuum diagrams. The momentum $p_i$ flows
      through the line $i$ as indicated by the arrow. 
      Each line may either be massless or carry mass $M$ and may be
    raised to an arbitrary integer power.
      }
  \end{center}
\end{figure}

In this paper we want to present the program package {\tt MATAD}
which was designed for the computation of
MAssive TADpoles at one-, two- and three-loop
order as pictured in Fig.~\ref{fig:tad123}. 
Thereby each line may be massless or carry the mass $M$.
In mathematical form the integrals to be solved by {\tt MATAD} read
\begin{eqnarray}
  &&
  \int\frac{{\rm d}^D p}{(2\pi)^D}
  \frac{1}{(p^2+M^2)^{n}}
  \,,
  \nonumber\\
  &&
  \int\frac{{\rm d}^D p}{(2\pi)^D}\frac{{\rm d}^D k}{(2\pi)^D}
  \frac{1}{(p_1^2+M_1^2)^{n_1}(p_2^2+M_2^2)^{n_2}(p_3^2+M_3^2)^{n_3}}
  \,,
  \nonumber\\
  &&
  \int\frac{{\rm d}^D p}{(2\pi)^D}\frac{{\rm d}^D k}{(2\pi)^D}
  \frac{{\rm d}^D l}{(2\pi)^D}
  \frac{1}{(p_1^2+M_1^2)^{n_1}(p_2^2+M_2^2)^{n_2}(p_3^2+M_3^2)^{n_3}} \times
  \nonumber\\&&\qquad\qquad\qquad\qquad\qquad\qquad
  \frac{1}{(p_4^2+M_4^2)^{n_4}(p_5^2+M_5^2)^{n_5}(p_6^2+M_6^2)^{n_6}}
  \,,
  \label{eq:inputints}
\end{eqnarray}
with $M_i=0$ or $M_i=M$. These expressions correspond to the diagrams in
Fig.~\ref{fig:tad123} where the momentum $p_i$ flows through the line $i$ as
indicated by the arrow. $p_i$ can be expressed as a linear combination of the
loop momenta. However, these relations are in our case not of interest.
Note that the integrals in Eq.~(\ref{eq:inputints}) are defined in Euclidean
space.

The key idea for the computation of tadpole integrals is based on the
integration-by-parts method~\cite{CheTka81} (see also
Appendix~\ref{app:ibp}). 
It can be used
for the derivation of recurrence relations which relate
vacuum integrals with different denominator structures.
The proper use of the recurrence relations allows the reduction of an
arbitrary 
integral to simple ones, which can be solved by successively using one- and
two-loop formulae, and a linear combination of a few so-called master
integrals. Only for 
them a hard calculation is necessary.
In the case of three-loop tadpole diagrams nine master integrals are needed.

At first sight the applications for vacuum integrals seem to be quite
restricted. However,
for diagrams involving several mass scales, which follow a certain
hierarchy, it is very often advantageous to apply
an asymptotic expansion~\cite{Smi95} which allows for a systematic expansion
in the inverse heavy scale.
Then the multi-scale integrals are expressed as
products of single scale ones. In~\cite{Har:diss} and~\cite{Sei:dipl}
the rules for the so-called large-momentum and 
hard-mass procedure have been
automated and computer programs, {\tt LMP}~\cite{Har:diss}
and {\tt EXP}~\cite{Sei:dipl}, have been developed. 
They generate for a given diagram all relevant sub-graphs together with the
administrative files which govern the very calculation.
{\tt LMP}~\cite{Har:diss}
is written in {\tt PERL} and can be applied to problems
where one large momentum is involved.
{\tt EXP}~\cite{Sei:dipl}, written in {\tt C++}, allows for
a successive use of the large-momentum and hard-mass procedure
and thus can deal with problems involving many scales.
Both programs produce output which can be read into
{\tt MATAD} and {\tt MINCER}.
Thus the combination of both massive vacuum integrals
and massless propagator-type diagrams is very powerful to attack
problems involving several different mass scales.
We want to mention that
{\tt MATAD} can be easily linked to a generator for Feynman diagrams.
More details --- in particular on the automation of the computation of Feynman
diagrams --- can be found in~\cite{HarSte98}.

The outline of the paper is as follows:
In Section~\ref{sec:structure} the structure of {\tt MATAD} and the
way it works is
described. With the help of this section the reader should be able to use
{\tt MATAD} for his own problems.
Deeper insight into some selected parts
is provided in Section~\ref{sec:details}.
In Section~\ref{sec:examples} explicit examples are discussed
and hints for the convenient usage of {\tt MATAD} are given.
In Appendix~\ref{app:ibp} the ideas of 
the integration-by-parts method are reviewed.
Appendix~\ref{app:topfile} lists all massive/massless combinations which are 
implemented into the topology files and in Appendix~\ref{app:not}
the notation of the input and output is described. Furthermore
the results for the master integrals are listed and the switches for the
input-file 
are described. Appendix~\ref{app:files} contains a list of all files
of {\tt MATAD} and, finally, in Appendix~\ref{app:testrun}
the complete output of one of the considered examples is listed. 


\section{Structure and mode of operation}
\label{sec:structure}

As {\tt MATAD} is completely written in {\tt FORM}~\cite{form} 
its installation
reduces to copying the individual files into the corresponding directories.
In the main directory the following files appear:

{\footnotesize
\begin{verbatim}
form.set  inc/  matadform  prc/  problems/
\end{verbatim}}
\noindent
The directories \verb|inc| and \verb|prc| contain the include-files and
procedures, respectively. They are described in more detail in
Section~\ref{sec:details} and Appendix~\ref{app:files}. 
\verb|matadform| is a shell script which calls {\tt FORM} in such a way
that files from sub-directories can be included. 
It has to be adjusted by the user by simply specifying the
corresponding paths.
The file \verb|form.set|
contains {\tt FORM}-specific settings which have to
be adjusted according to the
underlying platform. For details concerning the different switches we
refer to the {\tt FORM} manual~\cite{form}.
The user-specific files are all contained in the folder \verb|problems|.

There are at least two files which should be provided by the user:
{\tt main<prb>} and {\tt <prb>.dia} where {\tt <prb>} stands for the name of
the considered problem.
The first one contains apart from some parameters
essentially the information which diagram should be treated.
Some explicit examples are given below.
All the information about the diagrams, the projectors to be applied, etc.\
is contained in the file {\tt <prb>.dia}. It is built up by 
{\tt FORM} folds and splits into two parts so that its
generic structure looks as follows

{\footnotesize
\begin{verbatim}
*--#[ TREAT0:
[...]
*--#] TREAT0:

*--#[ TREAT1:
[...]
*--#] TREAT1:

*--#[ TREAT2:
[...]
*--#] TREAT2:

*--#[ TREATMAIN:
[...]
*--#] TREATMAIN:

*
* in the following list each diagram is contained in a separate FORM fold
*

*--#[ d1l1:
[... diagram 1 ...]
#define TOPOLOGY "XY"
*--#] d1l1:

*--#[ d1l2:
[... diagram 2 ...]
#define TOPOLOGY "XY"
*--#] d1l2:

[...]
\end{verbatim}}

\noindent
The first part consists of the first four folds --- the so-called special
treat files. They provide the
possibility to
interact at different stages and thus influence the computation.
Whereas {\tt TREAT0}, {\tt TREAT1} and {\tt TREAT2} are read before
the recurrence relations are applied the content of {\tt TREATMAIN}
is read right before the results are stored to disk.
The second part of {\tt <prb>.dia} contains a list of all diagrams to be
considered where each diagram is written in a separate fold. The name of these
folds is arbitrary. 

\begin{figure}[t]
  \begin{center}
    \epsfxsize=14cm
    \leavevmode
    \epsffile[60 200 550 730]{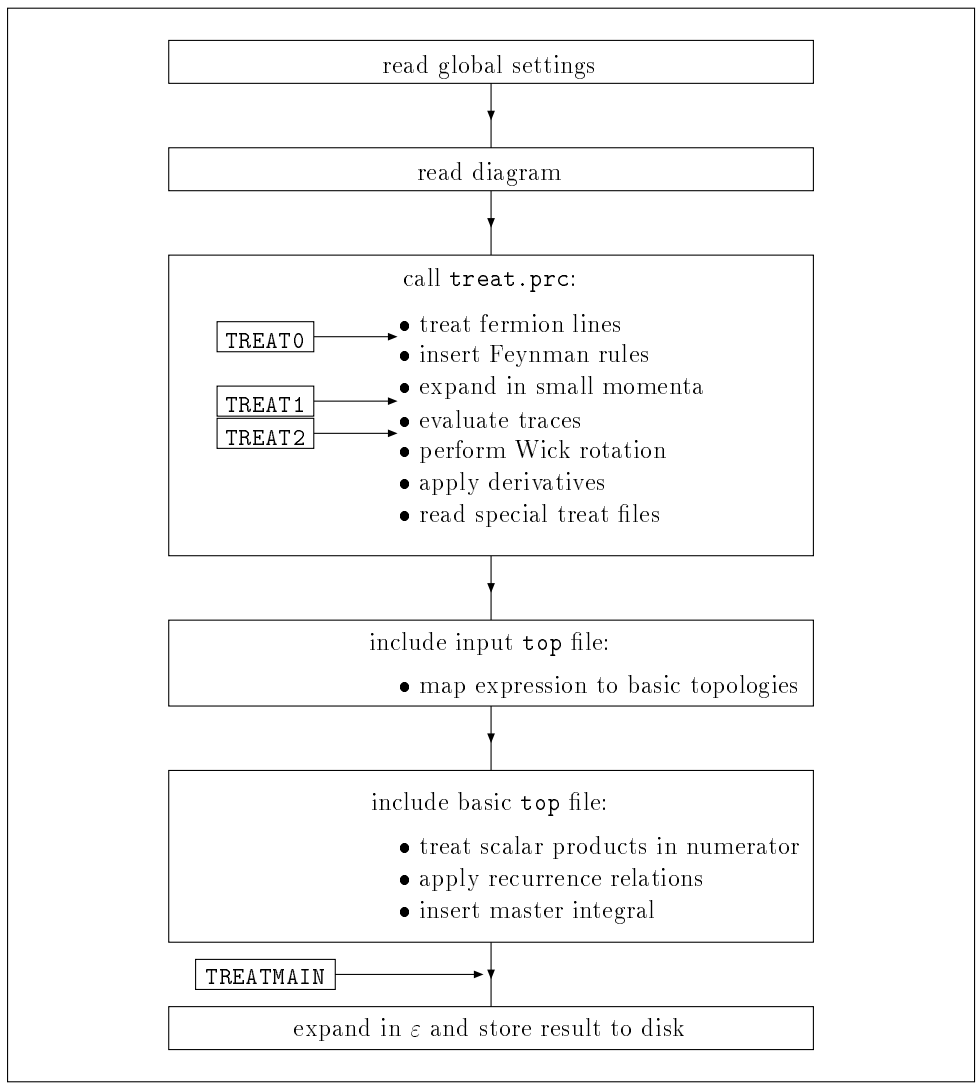}
    \caption{\label{fig:matad}Flowchart illustrating the structure
    of {\tt MATAD}.
      }
  \end{center}
\end{figure}

Once these two files are set up the calculation is simply initiated 
by calling the program {\tt main<prb>} and the following steps are
performed.
They are also illustrated in Fig.~\ref{fig:matad}.
\begin{enumerate}
\item
  Read global settings. They are partly contained in {\tt inc/main.gen}
  and should not be modified. Others can be set by the user
  in the file {\tt main<prb>}.
  They are described in Appendix~\ref{app:settings}.
\item
  Read the input data for the diagram specified in {\tt main<prb>}
  with the help of the variables {\tt PRB}, {\tt FOLDER} and {\tt DIAGRAM}.
  The generic {\tt FORM} command reads
  \verb|#include problems/'PRB'/'FOLDER'.dia # 'DIAGRAM'|.
\end{enumerate}
As a next step the file {\tt treat.prc} is called and the 
following operations are performed.

\begin{enumerate}
  \setcounter{enumi}{2}
\item
  Insert Feynman rules for functions appearing in the input.
  In a first step the fermions (encoded in the functions {\tt S}, {\tt SS},
  \ldots\,, cf.\ Appendix~\ref{app:notin}) are 
  resolved. It is important to do this before any contraction of indices is
  done. Then the propagators and vertices are treated.

  In the current version the QCD Feynman rules are implemented
  (except the four-gluon vertex; see
  Appendix~\ref{app:notin}). It is, however, straightforward to implement new
  vertices in the user-specific treat files.
\item
  Apply projector. This should be done in one of the
  special treat files. The optimal position depends on the integrals to be
  computed. From now on only scalar integrals without any free indices are
  present.
\item
  Expansion of the scalar denominators in the small quantities (mass and/or
  momentum).
\item
  Perform traces.
\item
  Do Wick rotation. This is done by multiplying each momentum
  by the imaginary unit (see also Appendix~\ref{app:not}).  
  From now on the expression is defined in Euclidean space.
\item
  Apply derivatives in order to factorize the external momentum.
  In this context see also the variables {\tt DALA12} and
  {\tt DALAQN} in Appendix~\ref{app:settings}.
\end{enumerate}
As there is the possibility to interact at three different places --- after
the fermions are treated ({\tt TREAT0}) and before and after the traces are
performed ({\tt TREAT1}, respectively, {\tt TREAT2})
the order of the commands may slightly be varied by the user.

At this stage
the scalar products in the numerator of the integrals
should be formed by either only loop momenta or only
external momenta (which then constitutes a trivial prefactor).
In the denominator the (scalar) propagators may be
raised to arbitrary power.

The next steps constitute the main part of {\tt MATAD}.
\begin{enumerate}
  \setcounter{enumi}{8}
\item
  Express the scalar products of the numerator in terms of the denominators.
  This produces a ``1'' in the numerator of the integrals.
  It might be that this step is very time and memory consuming.
\item
  Apply recurrence relations to reduce the number of different 
  integrals to simpler ones and 
  to a small set of master integrals.
\item
  Expand the result in $\varepsilon$
  and store it in the directory\\
  \verb|problems/'PRB'/results/'DIAGRAM'.res|
  under the name \verb|'DIAGRAM'|.
\end{enumerate}
An expansion in $\varepsilon$ is also done at various intermediate steps.
Although poles of at most third order may appear for a three-loop vacuum
integral terms up to order $\varepsilon^6$ have to be kept in the
expansion as artificial poles may appear during the application of the
recurrence relations 
(cf.\ step~10).

\begin{figure}[t]
  \begin{center}
    \epsfxsize=14cm
    \leavevmode
    \epsffile[53 350 540 730]{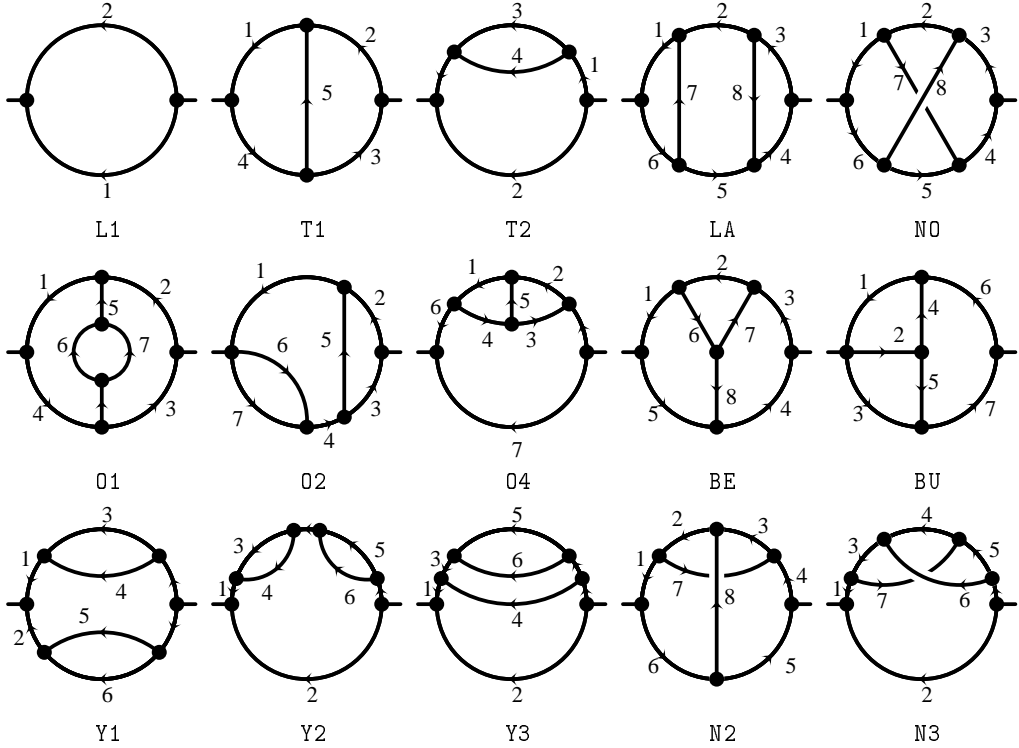}
    \caption{\label{fig:intop}
    The input of the diagram to be computed can be mapped to one of
    these topologies. The implemented massive/massless combinations can
    be found in Appendix~\ref{app:topfile}.
    The momentum $p_i$ flows through line $i$ as indicated by the arrow.
      }
  \end{center}
\end{figure}

Steps~9 and~10 constitute the central part
of {\tt MATAD}. They heavily depend on the loop-order and the 
topology which has
to be specified apart from the very diagram in the folds
{\tt d1l1}, {\tt d1l2}, \ldots (see above). Thus let us elaborate on this
point in the following. In principle 
it suffices to define one input topology at one-,
two- and three-loop order, where the number of internal lines amount to 
one, three and six, respectively (see Fig.~\ref{fig:tad123}).
If one allows each line to be massless and carry the mass $M$ at the same time
these three topologies are sufficient to cover all possible cases that can
occur in the calculation of one-, two- and three-loop vacuum integrals.
Note that a partial fractioning for terms like
\begin{eqnarray}
  \frac{1}{\left(p^2\right)^a\left(p^2+M^2\right)^b}
  \,,
\end{eqnarray}
where $a$ and $b$ are positive integers, leads to the same topologies
with the only difference that now each line is either massless of massive.
It is, however, not at all practical to rewrite the input to the notation of
Fig.~\ref{fig:tad123} before generating the file {\tt dia.<prb>}.
On the contrary it is advantageous to enlarge the input topologies.
Currently the topologies shown in Fig.~\ref{fig:intop} are implemented in
{\tt MATAD}.
The momentum $p_i$ flowing through line $i$ can be expressed as a linear
combination of the loop momenta. For our purpose these relations are, however,
not of interest.
The choice of the topologies
was guided by the package {\tt MINCER}~\cite{mincer}
and for convenience
the same notation concerning the definition of the momenta $p_i$ has been
adopted.
The implemented massive/massless
combinations of each topology are listed in Appendix~\ref{app:topfile}.
After the declaration of the diagram in the folds
{\tt d1l1}, {\tt d1l2}, \ldots the corresponding topology is specified 
via

{\footnotesize
\begin{verbatim}
  #define TOPOLOGY "XY"
\end{verbatim}}

\noindent
where \verb|XY| corresponds to one of the topologies of Fig.~\ref{fig:intop}.

\begin{figure}[t]
  \begin{center}
    \epsfxsize=14cm
    \leavevmode
    \epsffile[80 430 500 730]{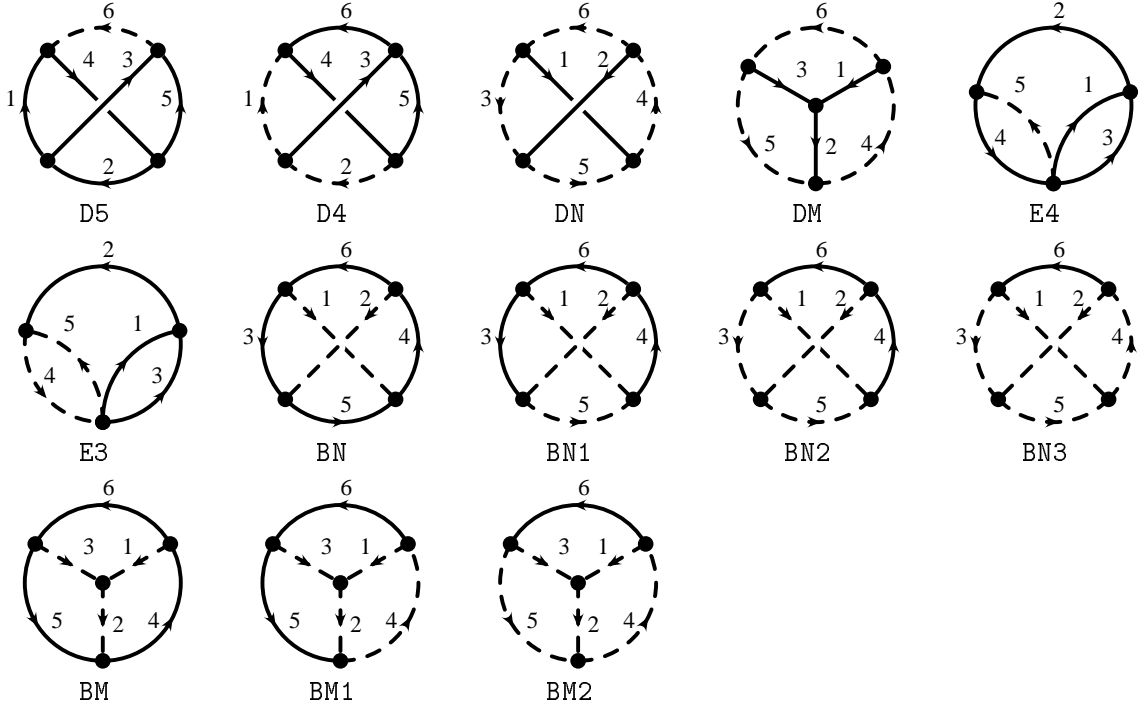}
    \caption{\label{fig:batop}
      Basic three-loop topologies implemented into {\tt MATAD}.
      They are reduced to either simple integrals or master 
      integrals with the help of recurrence relations. 
      }
  \end{center}
\end{figure}

The notation concerning the momenta as introduced in Fig.~\ref{fig:intop} is
quite convenient to be used for the input. However, the 
very recursion procedure is formulated for the three-loop topology
of Fig.~\ref{fig:tad123} where the lines are either massive or massless.
This leads to $14$ different cases which are classified in~\cite{Avd97}.
Thus before step~9 is performed the momenta 
are transformed from the notation of Fig.~\ref{fig:intop}, which is used 
in the input, to the so-called basic topologies shown in
Fig.~\ref{fig:batop}. 
For them --- after decomposing the scalar products in the numerator into parts
of the denominator --- the very recursion procedure is performed.
Note that at this stage all propagators may be raised to an arbitrary
integer power.

\begin{figure}[t]
  \begin{center}
    \epsfxsize=5cm
    \leavevmode
    \epsffile[189 293 437 499]{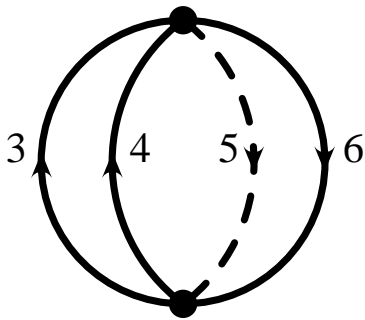}
    \caption{\label{fig:master_add}
      Master diagram resulting from topology
      {\tt BN1}. Here all propagators are raised to power one.
      Its constant part contains the expression {\tt S2} and its
      ${\cal O}(\varepsilon)$ part {\tt OepS2}
      as listed in Appendix~\ref{app:master}.
      }
  \end{center}
\end{figure}

Actually some of the three-loop diagrams (e.g.\ \verb|BN3|)
can be computed by the successive use of one- and two-loop
procedures for massless propagator type diagrams or vacuum integrals,
respectively. In such cases some of the (one- and two-loop)
routines from {\tt MINCER}~\cite{mincer} are used for parts of
the computation.
The corresponding procedures are listed in Appendix~\ref{app:files}.
For other cases (e.g.\ \verb|E4| or \verb|BN2|)
simple relations reduce one of the lines to 
zero and the resulting diagram can again be solved easily.
Only for the cases \verb|D5, D4, DN, DM, E3, BN| and \verb|BN1|
the recursion procedure has to be applied
until one arrives at master integrals.
They coincide with the corresponding
diagrams of Fig.~\ref{fig:batop} where all denominators are raised to
power one.
Only the one pictured in Fig.~\ref{fig:master_add}, which results from
\verb|BN1|, is needed in addition. 
From this diagram even the ${\cal O}(\varepsilon)$ part is required.
The analytic expressions are 
given in Appendix~\ref{app:master}.
It should be mentioned 
that the recurrence relations for \verb|BM| are quite involved. However,
for this topology no difficult master integral is needed.
Note that the basic topologies \verb|BM1| and \verb|BM2| actually
coincide with \verb|BN2| and \verb|BN3|, respectively. For convenience
partly different codes had been written which actually
turned out to be quite useful while debugging {\tt MATAD}.

The recurrence relations leading to simple diagrams and master integrals for
the three-loop topologies are derived in Refs.~\cite{Bro92}
and~\cite{Avd97}. Except the case where all six lines are massive
all of them are implemented into {\tt MATAD}.
The reason why this case is missing is simply that it was not yet needed for
practical calculations. All master integrals are listed in
Appendix~\ref{app:master}. 

\begin{figure}[t]
  \begin{center}
    \epsfxsize=14cm
    \leavevmode
    \epsffile[70 600 550 700]{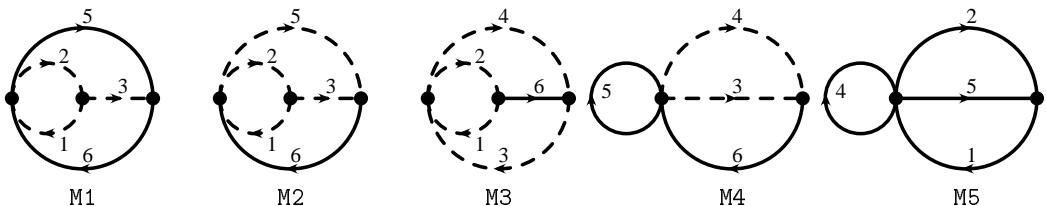}
    \caption{\label{fig:simple}
      Simple integrals which occur as a result of the recurrence
      procedure next to the master integrals. All propagrators may be
    raised to an arbitrary power.
      }
  \end{center}
\end{figure}

In Fig.~\ref{fig:simple} the simple integrals are listed which also result
from the recursion procedure. They can all be expressed in terms of
$\Gamma$ functions by the successive use of results for massless
one-loop two-point functions, $P_{ab}(Q)$, in combination with one-
($V_a$) and two-loop ($V_{abc}$) vacuum integrals. 
For convenience we want to list the explicit results for 
$P_{ab}(q)$, $V_a$ and $V_{abc}$ in Euclidean space:
\begin{eqnarray}
P_{ab}(Q) &=& \int\frac{{\rm d}^Dp}{\left(2\pi\right)^D}
              \frac{1}{p^{2a}\left(p+Q\right)^{2b}}
\nonumber\\
          &=& 
\frac{\left(Q^2\right)^{D/2-a-b}}{\left(4\pi\right)^{D/2}}
\frac{
  \Gamma(a+b-D/2)
  \Gamma(D/2-a)
  \Gamma(D/2-b)
}{
  \Gamma(a)
  \Gamma(b)
  \Gamma(D-a-b)
}  
\,,
\nonumber\\
V_{a}&=& \int\frac{{\rm d}^Dp}{\left(2\pi\right)^D}
              \frac{1}{\left(p^2+M^2\right)^{a}}
\,\,=\,\,
\frac{\left(M^2\right)^{D/2-a}}{\left(4\pi\right)^{D/2}}
\frac{
  \Gamma(a-D/2)
}{
  \Gamma(a)
}  
\,,
\nonumber\\
V_{abc} &=&
  \int\frac{{\rm d}^D p}{(2\pi)^D}\frac{{\rm d}^D k}{(2\pi)^D}
  \frac{1}{(p^2+M^2)^{a}(k^2+M^2)^{b}(\left(p+k\right)^2)^{c}}
\nonumber\\
          &=& 
\frac{\left(M^2\right)^{D-a-b-c}}{\left(4\pi\right)^{D}}
\frac{
  \Gamma(a+b+c-D)
  \Gamma(a+c-D/2)
  \Gamma(b+c-D/2)
  \Gamma(D/2-c)
}{
  \Gamma(a)
  \Gamma(b)
  \Gamma(a+b+2c-D)
  \Gamma(D/2)
}
\,.
\nonumber\\
\label{eq:simpint}
\end{eqnarray}

\begin{figure}[ht]
  \begin{center}
  \leavevmode
      \epsfxsize=5cm
      \epsffile[131 241 481 560]{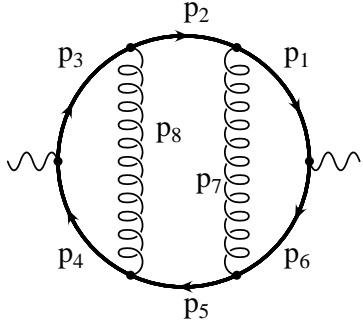}\\
      \caption[]{\label{fig:ladder}
        Three-loop diagram of type {\tt LA}.
        }
  \end{center}
\end{figure}

The vacuum
integrals occurring at one- and two-loop level are quite simple. Actually
most of them can be expressed in terms of $\Gamma$ functions for arbitrary
exponents of the propagators and no recursion relations are 
needed (cf.\ Eq.~(\ref{eq:simpint})).
Only for the two-loop integral in Fig.~\ref{fig:tad123} where all three lines 
carry the mass $M$ it is useful to implement simple recurrence relations which
reduce the integrals to one master integral 
where all exponents are raised to the first power
only. For three-loop calculations the result of this integral is
needed up to ${\cal O}(\varepsilon)$ 
({\tt T1ep}, see Appendix~\ref{app:master}).

For definiteness we want to consider an explicit example.
Let us consider the three-loop diagram of Fig.~\ref{fig:ladder}
which we would like to expand up to fourth order
in the external momentum, $q_1$. For simplicity we neglect the tensor
structure and consider only the scalar integral obtained in the case when
the full line represents a massive particle and the curly line a massless
one.
In the directory {\tt problems/scalar/} one can find the following files

{\footnotesize
\begin{verbatim}
mainscalar  results/  scalar.dia
\end{verbatim}}

\noindent
where {\tt results/} is a directory to store the result of the diagram.
The file {\tt scalar.dia} looks as follows
{\footnotesize

\begin{verbatim}
*--#[ TREAT0:
*--#] TREAT0:

*--#[ TREAT1:
*--#] TREAT1:

*--#[ TREAT2:
*--#] TREAT2:

*--#[ TREATMAIN:
*--#] TREATMAIN:

*--#[ scalar:
        M^4
        *s1m
        *s2m
        *s3m
        *Dh(p4,q1)
        *Dh(p5,q1)
        *Dh(p6,q1)
        /p7.p7
        /p8.p8
        ;

        #define TOPOLOGY "LA"
*--#] scalar:
\end{verbatim}}

\noindent
In this case no special treat file is needed.
The very diagram can be found in the fold {\tt scalar} where the
multiplication with $M^4$
is done in order to end up with a dimensionless expression.
The external momentum is routed through the lines 4, 5 and 6 as can be seen in
the arguments of the function \verb|Dh|. The other massive propagators are
denoted by \verb|s1m|, \verb|s2m| and \verb|s3m|. The massless lines translate
into the factors \verb|1/p7.p7| and \verb|1/p8.p8|.
For more details on the notation we refer to Appendix~\ref{app:notin}.
The file {\tt mainscalar} looks as follows:

{\footnotesize
\begin{verbatim}
#define PRB "scalar"
#define DALAQN "q1"
#define GAUGE "0"
#define POWER "4"
#define CUT "0"
#define FOLDER "scalar"
#define DIAGRAM "scalar"
#-
#include main.gen
\end{verbatim}}

\noindent
With the command 

{\footnotesize
\begin{verbatim}
#define POWER "4"
\end{verbatim}}

\noindent
we require an expansion up to fourth order in $q_1$ and

{\footnotesize
\begin{verbatim}
#define DALAQN "q1"
\end{verbatim}}

\noindent
effectively factors out the terms $(q_1^2)^n$.
The calculation is initiated with the command

{\footnotesize
\begin{verbatim}
> matadform problems/scalar/mainscalar
\end{verbatim}}

\noindent
and after a few seconds the final result is displayed on the screen

{\footnotesize
\begin{verbatim}
   scalar =
       - 2 + 2*z3 - Q1.Q1*z3*M^-2 + 227/216*Q1.Q1*M^-2 + 1/2*Q1.Q1^2*z3*M^-4
          - 1876/3375*Q1.Q1^2*M^-4;
\end{verbatim}}

\noindent
As expected it is finite and contains three terms in the expansion in
$q_1^2/M^2$.
The result is stored to the directory {\tt problems/scalar/results/} and can
be read with the {\tt FORM} command {\tt load}.
Details on the notation of the output and the used conventions
can be found in Appendix~\ref{app:not}.


\section{Some details on the internal structure}
\label{sec:details}

Once the recurrence relations are implemented it is in principle
possible to compute diagrams of arbitrary complexity. However, in practice
one arrives quite soon at the limits set by the soft- or hardware.
The algebra language {\tt FORM} is designed to deal with large
expressions. Still it happens quite easily that in internal steps the
expressions exceed several Megabytes and even approach the order of a few
Gigabyte. This significantly slows down the performance and it is advantageous
to implement several tricks. Some of them are described in this section.

$\bullet$
During the recurrence procedure it happens very often that whole blocks
of commands have to be executed until the recursion has reached an end. Within
{\tt FORM} there are the commands

{\footnotesize
\begin{verbatim}
  repeat;
    [...]
  endrepeat;
\end{verbatim}}

\noindent
which in principle allows for such a construction. However, in practice it is
not possible to use between \verb|repeat| and \verb|endrepeat| a command which
forces {\tt FORM} to combine identical terms like, e.g., \verb|.sort|.
For this reason a preprocessor variable, \verb|NOR| (see also
Appendix~\ref{app:settings}), has been introduced
which in combination with the construction

{\footnotesize
\begin{verbatim}
  #do i=1,'NOR'
    [...]
  #enddo
\end{verbatim}}

\noindent
allows the use of \verb|.sort| in intermediate steps of the recursion
commands. 

$\bullet$
At this point we should also mention that the procedure \verb|ACCU| has been
adopted from {\tt MINCER}~\cite{mincer}. It collects in the 
argument of the function \verb|acc()| the
polynomials in $\varepsilon$ and thus significantly reduces the number
of terms. E.g., the expression

{\footnotesize
\begin{verbatim}
  x1*x2 + 4*ep*x1*x2 + 12*ep^3*x1*x2
\end{verbatim}}

\noindent
transforms after \verb|#call ACCU{test}| to

{\footnotesize
\begin{verbatim}
  acc(1 + 4*ep + 12*ep^3)*x1*x2
\end{verbatim}}

\noindent
and instead of three only one term has to be treated in the following
commands.

$\bullet$
Very often it happens that propagators of the type 
$1/(M^2-(p+q)^2)^n$ have to be expanded in the momentum $q$
and afterwards derivatives w.r.t. $q$ are
applied in order to factor out 
powers of $q^2$. It turns out that it is very useful to
expand in a first step only the part $2pq$ and keep the factors $q^2$
unexpanded in the form $1/(M^2-(p^2+q^2))^n$. 
Thus less terms have to be considered while the derivatives are applied.
Afterwards the expansion in $q^2$ is performed.
A related discussion can also be found in~\cite{Tar95}.

$\bullet$
The expansion of the scalar denominators in a small momentum is also
very time consuming --- especially for high values of '{\tt POWER}'.
In order to do this in an effective way the variable {\tt poco}, which
is an abbreviation for ``power counting'', is defined 
via

\noindent
{\footnotesize
\begin{verbatim}
S poco(:'POWER');
\end{verbatim}}

\noindent
after the declaration of the preprocessor variable {\tt POWER}.
This definition ensures that the terms involving {\tt poco$^n$}
with $n>$\verb|'POWER'| are automatically set to zero.
Depending on the problem {\tt poco} should also be
considered in the special treat files.

$\bullet$
The application of the recurrence relations can lead to 
spurious $1/\varepsilon$ poles (cf.\ Appendix~\ref{app:ibp}) which are
in general quite dangerous if an expansion in $\varepsilon$ is
performed in intermediate steps.
In {\tt MATAD} at most three $1/\varepsilon$ poles 
arise from the recurrence relations. Together with a possible 
$1/\varepsilon^3$ term from the three-loop integrals an expansion up
to order $\varepsilon^6$ has to be done in intermediate steps in order
to get the correct constant term.

$\bullet$
For quite a lot of applications the topology \verb|BN|
(cf.\ Fig~\ref{fig:batop})
plays a crucial role. The original recurrence procedure for this toplogy to
master integrals was proposed in~\cite{Bro92}: In a first step the
exponent of three out of the four massive denominators are reduced to
one. Then the exponents of the massless lines are reduced to zero. Finally the
remaining 
line is treated and one arrives (apart from simple integrals) at an
integral consisting of four massive lines connecting two vertices.
It is connected to the corresponding master 
integral of Fig.~\ref{fig:batop} through a simple relation.

The equation involved in the last recursion
step is quite involved. It generates from each term more than ten terms 
at each call.
Note that the exponent of the last massive denominator gets increased by the
proceeding steps.
This enormously slows down the calculation --- in some cases it makes it even
impossible. The idea to circumvent this problem is based on the observation 
that
the massless exponents can be reduced to zero even if none of the massive ones
is reduced to one.
The corresponding recurrence relations are short and thus one arrives 
with only little effort at three-loop integrals with four massive lines,
$B_N(0,0,n_3,n_4,n_5,n_6)$ (cf.\ Fig.~\ref{fig:master_add} where all lines are
massive and the exponents of the denominators are given by $n_3$,
$n_4$, $n_5$ and $n_6$).
These diagrams are now treated in the following way:
Temporarily an external momentum is introduced which flows through one of the
lines. We choose line 3 for definiteness.
In a second step the operator 
$\Box_q=\partial/\partial q^\mu \partial/\partial q_\mu$ is applied and $q$ is
set to zero afterwards. This leads to an equation connecting 
$B_N(0,0,n_3,n_4,n_5,n_6)$,
$B_N(0,0,n_3-1,n_4,n_5,n_6)$ and
$B_N(0,0,n_3-2,n_4,n_5,n_6)$
\begin{eqnarray}
  B_N(0,0,n_3,n_4,n_5,n_6) &=&
  -\frac{1}{4M^2(n_3-2)(n_3-1)} \Box_q B_N(0,0,n_3-2,n_4,n_5,n_6)
  \nonumber\\&&\mbox{}
  +\frac{-2D+4(n_3-1)}{4M^2(n_3-1)} B_N(0,0,n_3-1,n_4,n_5,n_6)
  \,,
  \label{eq:BNrec}
\end{eqnarray}
which can be applied until $n_3=3$. As the other indices are not affected the
same procedure can be applied to them as well and one ends up with the
integrals
$B_N(0,0,1,1,1,1)$,
$B_N(0,0,2,1,1,1)$,
$B_N(0,0,2,2,1,1)$,
$B_N(0,0,2,2,2,1)$ and
$B_N(0,0,2,2,2,2)$.
Both their values for $q=0$ and the result for the application of 
$(\Box_q)^n$ has to be known where the index $n$ depends on 
how often Eq.~(\ref{eq:BNrec}) had to be applied.
The overall number of the integrals needed is still small and for 
most practical applications well below 100. They can be computed once and for
all using the method of Ref.~\cite{Bro92} and can be collected in a table.
Currently the table contains all results up to $n=10$
and partly for $n=11$.
In case the table is too small the original recurrence
procedure~\cite{Bro92} has to be used. This is done by defining the
preprocessor variable {\tt BNRECOLD} in the main file.


\section{Examples}
\label{sec:examples}

In this section we want to discuss some typical examples which can be
treated with {\tt MATAD}. In particular the content of the {\tt TREAT} folds
and the switches in {\tt main.<prb>} shall be discussed in detail.
In Section~\ref{sec:photon} two-loop corrections to the photon polarization
function are considered and in Section~\ref{sec:hgg} 
the calculation of three-loop vertex
corrections contributing to the Higgs decay into gluons is discussed.
As a last example we consider the computation of the fermion propagator 
in the limit of a small external momentum. 

\subsection{\label{sec:photon}Photon polarization function}

In this section only a calculation of two-loop diagrams is presented.
However, we want to take this opportunity to show a concept which can also be
used for the treatment of more complex problems.
In particular it is shown how the complete calculation can be automated
and one can ensure that all results are indeed up-to-date.

To be precise, we want to consider the two-loop diagrams induced by a
massive quark to the photon polarization function.
The problem shall be called {\tt Pi}. Then the file {\tt Pi.dia} looks as
follows:

{\footnotesize
\begin{verbatim}
*
* problem: Pi
*

*--#[ TREAT0:
  #message project out transversal part
  multiply, (d_(mu1,mu2)-q1(mu1)*q1(mu2)/q1.q1)*deno(3,-2);
  .sort
*--#] TREAT0:

*--#[ TREAT1:
*--#] TREAT1:

*--#[ TREAT2:
*--#] TREAT2:

*--#[ TREATMAIN:
*--#] TREATMAIN:

*
* 2-loop diagrams for problem Pi
*

*--#[ d2l1:
	((-1)
	*Dg(nu1,nu2,p5)
	*S(mu1,q1,p1m,nu1,q1,p2m,mu2,p3m,nu2,p4m)
	*1);

	#define TOPOLOGY "T1"
*--#] d2l1:

*--#[ d2l2:
	((-1)
	*Dg(nu1,nu2,p4)
	*S(mu1,q1,-p2m,mu2,p1m,nu2,p3m,nu1,p1m)
	*1);

	#define TOPOLOGY "T2"
*--#] d2l2:

*--#[ d2l3:
	((-1)
	*Dg(nu1,nu2,p4)
	*S(nu1,p3m,nu2,p1m,mu2,-q1,-p2m,mu1,p1m)
	*1);

	#define TOPOLOGY "T2"
*--#] d2l3:
\end{verbatim}}

\noindent
The function {\tt deno(x,y)} means $1/(x+y\varepsilon)$
and can be used for denominators.
{\tt TREAT0} contains the projector to the transversal part and
the three contributing diagrams are named
{\tt d2l1}, {\tt d2l2} and {\tt d2l3}.
The external momentum is denoted by \verb|q1|. The mass of the
quarks, which in the final result appears as {\tt M}, is introduced
through adding the symbol {\tt m} to the corresponding momentum
which is defined through the topologies {\tt T1} and {\tt T2} in
Fig.~\ref{fig:intop}.
We want to assume that $q_1^2\ll M^2$ and thus perform an expansion in 
$q_1$. This is achieved with the notation
\verb|S(...,q1,p1m,...)|. For more details concerning the notion we
refer to Appendix~\ref{app:not}.

The file {\tt mainPi} reads:

{\footnotesize
\begin{verbatim}
#define PRB "Pi"
#define PROBLEM0 "1"
#define DALAQN "q1"
#define GAUGE "xi"
#define POWER "4"
#define CUT "1"
#define FOLDER "Pi"
***#define DIAGRAM "d2l1"
#-
#include main.gen
\end{verbatim}}

\noindent
For most of the specified variables we refer to
Appendix~\ref{app:settings}. We only want to mention that 
a general gauge parameter \verb|xi| is chosen with the command
\verb|#define GAUGE "xi"|. This allows for an explicit check that the final
result, i.e. the sum of the three diagrams is gauge parameter independent.
The definition \verb|#define POWER "4"| requests for an expansion 
up to fourth order in \verb|q1|. The definition of the variable \verb|DIAGRAM|
is commented as it will be defined during the call of {\tt mainPi} (see
below). 

The computation of each diagram could be started separately and the
results could be summed at the end. Instead we want to take the opportunity to 
present a method which is unavoidable for problems where a large
number of diagrams contribute. 
In the following we want to present two more files which can easily be adopted
to other problems. The first one, {\tt makePi}, is a so-called {\tt
GNU make} file
and could look as follows:

{\footnotesize
\begin{verbatim}
SHELL = /bin/sh

DIA2 = \
      problems/Pi/results/d2l1.res\
      problems/Pi/results/d2l2.res\
      problems/Pi/results/d2l3.res

problems/Pi/results/Pi.2.res: $(DIA2)
        matadform problems/Pi/comPi > problems/Pi/log/comPi.log

ii = $(notdir $(basename $@))

$(DIA2): problems/Pi/mainPi
        if [ -f problems/Pi/results/$(ii).res ]; \
           then rm problems/Pi/results/$(ii).res; fi
        time matadform -d DIAGRAM=$(ii) \
          problems/Pi/mainPi > problems/Pi/log/$(ii).log; 
        if [ -f problems/Pi/results/$(ii).res ]; \
           then rm problems/Pi/log/$(ii).log; fi
\end{verbatim}}

\noindent
For details concerning the individual commands we refer to the
literature~\cite{make}. For us it is only important that
at the beginning the diagrams we want to compute are listed. Furthermore,
after the line
`\verb|problems/Pi/results/Pi.2.res: $(DIA2)|'
the command is given which specifies what shall be done once the computation
of the individual diagrams is finished: The diagrams are summed with
the help of the program {\tt comPi}

{\footnotesize
\begin{verbatim}
*
* comPi
*
#-
#include declare.matad
#define PRB "Pi"
.global

#do i=1,3
  load problems/'PRB'/results/d2l'i'.res;
#enddo

g res'PRB'2 = 
#do i=1,3
  + d2l'i'
#enddo
;

b ep;
print;
.store

#+
save problems/'PRB'/results/res'PRB'.2.res res'PRB'2;
.end
\end{verbatim}}

\noindent
To initiate the calculation one simply has to specify the command

{\footnotesize
\begin{verbatim}
> make -f problems/Pi/makePi
\end{verbatim}}

\noindent
where {\tt make} has to call the 
{\tt GNU} version~\cite{make} of the {\tt make} command.
The final result is stored in the file
{\tt problems/Pi/results/resPi.2.res}.
It reads

{\footnotesize
\begin{verbatim}
   resPi2 =
       + ep^-1 * (  - 6*Q1.Q1 + 8/5*Q1.Q1^2*M^-2 )

       + ep * (  - 35/6*Q1.Q1 - 6*Q1.Q1*z2 + 8/5*Q1.Q1^2*M^-2*z2 + 3116/1215*
         Q1.Q1^2*M^-2 )

       + 13/3*Q1.Q1 - 128/405*Q1.Q1^2*M^-2;
\end{verbatim}}

\noindent
Indeed, as expected, there is no trace of the gauge parameter
\verb|xi|.

\subsection{\label{sec:hgg}Higgs decay into two gluons}

Party for convenience and partly for historical reasons the notation
of the input topologies in Fig.~\ref{fig:intop} is closely connected
to two-point functions. However, {\tt MATAD} only deals with vacuum
diagrams independent of the number of external legs.
In this section we want to show that also problems involving at first sight
three-point functions can be
approached using {\tt MATAD}. 

\begin{figure}[ht]
  \begin{center}
  \leavevmode
      \epsfxsize=5cm
      \epsffile[189 314 481 478]{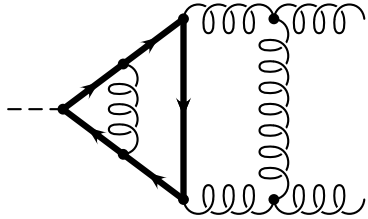}\\
      \caption[]{\label{fig:hgg}Sample diagram contributing to the
  decay of the Higgs boson. Solid and looped lines represent quarks
  and gluons, respectively.
        }
  \end{center}
\end{figure}

Let us consider QCD corrections to the decay of the
Standard Model Higgs boson
into two gluons. It is convenient to construct an effective theory where the
top quark is integrated out. Details on the theoretical background
can be found in~\cite{CheKniSte97hgg}.
Here it shall only be mentioned that the coefficient function which contains
the dependence on the mass of the top quark, $M_t$,
can be computed from triangle diagrams as pictured in 
Fig.~\ref{fig:hgg}.
According to their Lorentz structure the result can be written as follows
\begin{eqnarray}
  K(M_t) \, \left( q_1^\nu q_2^\mu - q_1 q_2 g^{\mu\nu} \right)
  \,,
  \label{eq:c1}
\end{eqnarray}
where $q_1$ and $q_2$ are the momenta of the gluons with polarization
vectors $\epsilon^\mu(q_1)$ and $\epsilon^\nu(q_2)$.
Thus the vertex diagrams have to be expanded up to linear order both
in $q_1$ and $q_2$ and an appropriate projector has to be applied 
in order to get $K(M_t)$.

The file containing the diagrams could look as follows

{\footnotesize
\begin{verbatim}
*
* problem: hgg
*

*--#[ TREAT0:
multiply, (
  a*deno(2,-2)*(q1.q2*d_(mu,nu)-q2(nu)*q1(mu)-q2(mu)*q1(nu))
 +b*deno(2,-2)*(-q1.q2*d_(mu,nu)+(3-2*ep)*q2(nu)*q1(mu)+q2(mu)*q1(nu))
);
.sort
*--#] TREAT0:

*--#[ TREAT1:
*--#] TREAT1:

*--#[ TREAT2:
*--#] TREAT2:

*--#[ TREATMAIN:
*--#] TREATMAIN:

*--#[ d3l335:
        ((-1)
        *M
        *Dg(nu1,nu2,p1)
        *Dg(nu7,nu8,-p4)
        *Dg(nu3,nu4,q1,-p1)
        *Dg(nu5,nu6,q2,p1)
        *S(-q1,-p3m,nu7,-q1,p5m,nu4,-p2m,nu6,q2,p5m,nu8,q2,-p3m)
        *V3g(mu,q1,nu1,-p1,nu3,p1-q1)
        *V3g(nu,q2,nu2,p1,nu5,-p1-q2)
        *1);

        #define TOPOLOGY "O4"
*--#] d3l335:
\end{verbatim}}

\noindent
where the diagram {\tt d3l335} corresponds to the one shown in
Fig.~\ref{fig:hgg}.
The fold {\tt TREAT0} contains (up to an overall factor
$(q_1.q_2)^{-2}$) the projector on the coefficients in front of
the structures $g^{\mu\nu}$ and $q^\nu_1 q^\mu_2$ of Eq.~(\ref{eq:c1}).
They are marked by 
the symbols \verb|a| and \verb|b|, respectively.
Thus the transversality of Eq.~(\ref{eq:c1}) can be explicitly checked
in the sum of all contributing diagrams (the result of a single
diagram does in general not have a transverse structure).
 
The corresponding {\tt main}-file is very similar to the one listed
in Section~\ref{sec:photon}. The only difference (apart from replacing
{\tt Pi} by {\tt hgg}) is that the commands

{\footnotesize
\begin{verbatim}
#define DALAQN "q1"
#define GAUGE "xi"
#define POWER "4"
#define CUT "1"
\end{verbatim}}

\noindent
should be replaced by

{\footnotesize
\begin{verbatim}
#define DALA12 "1"
#define GAUGE "0"
#define POWER "2"
#define CUT "0"
\end{verbatim}}

\noindent
The third line
ensures that an expansion of the integrand up to second order in the 
external momenta is performed and the first one sets
$q_1^2$ and $q_2^2$ to zero and factors out
the scalar product $q_1 q_2$.
\verb|#define CUT "0"| sets $\varepsilon$ to zero in the final result
as the terms of ${\cal O}(\varepsilon)$
are anyway not computed completely at three-loop order.
In this example we choose Feynman gauge which is achieved with 
\verb|#define GAUGE "0"|.

After calling {\tt MATAD} it takes of the order of a minute 
to obtain the result:

{\footnotesize
\begin{verbatim}
   d3l335 =
       + ep^-2 * ( 40*Q1.Q2*M^2*a + 344/9*Q1.Q2^2*a - 232/9*Q1.Q2^2*b )

       + ep^-1 * (  - 308/3*Q1.Q2*M^2*a - 3530/27*Q1.Q2^2*a + 1786/27*Q1.Q2^2*
         b )

       + 60*Q1.Q2*M^2*z2*a + 734/3*Q1.Q2*M^2*a - 1936/9*Q1.Q2^2*z3*a + 1136/9*
         Q1.Q2^2*z3*b + 172/3*Q1.Q2^2*z2*a - 116/3*Q1.Q2^2*z2*b + 46817/81*
         Q1.Q2^2*a - 26239/81*Q1.Q2^2*b;
\end{verbatim}}

\noindent
Note that the terms proportional to \verb|Q1.Q2*M^2| cancel after
adding all contributing diagrams.

\subsection{\label{sub:ferm}Fermion propagator}

In this example we compute the small-momentum expansion
of a three-loop diagram which contributes to the
fermion propagator. Recently these kind of diagrams have been considered in 
different kinematical regions in order to obtain the three-loop
relation between the $\overline{\rm MS}$ and on-shell quark
mass~\cite{CheSte99,CheSte00}. 
This subsection contains all relevant input information
whereas the complete output is given in Appendix~\ref{app:testrun}.

The fermion self energy, $\Sigma(q)$,
can be decomposed into a scalar and vector part
\begin{eqnarray}
  \Sigma(q) &=& M \Sigma_S(q^2) + q\hspace{-.45em}/\,\Sigma_V(q^2) 
  \,,
  \label{eq:sigma}
\end{eqnarray}
where $q$ is the external momentum and $M$ is the mass of the quark. 
We are interested in the computation of the scalar functions 
$\Sigma_S$ and $\Sigma_V$.

The file \verb|mainfp| looks as follows

{\footnotesize
\begin{verbatim}
#define PRB "fp"
#define PROBLEM0 "1"
#define DALAQN "q1"
#define GAUGE "0"
#define POWER "1"
#define CUT "0"
#define FOLDER "fp"
#define DIAGRAM "d3l79"
#-
#include main.gen
\end{verbatim}}

\noindent
The variable \verb|POWER| is defined in such a way that an expansion up to
third order is performed. This leads to the constant and order $q^2/M^2$ terms
for the functions $\Sigma_S$ and $\Sigma_V$ as $\Sigma(q)$ itself
has mass dimension one.
The corresponding projectors and the diagram to be computed look as follows
(\verb|fp.dia|):

{\footnotesize
\begin{verbatim}
*
* problem: fp
*

*--#[ TREAT0:
  multiply, 1/4*(1/M + a * g_(1,q1)/q1.q1);
*--#] TREAT0:

*--#[ TREAT1:
*--#] TREAT1:

*--#[ TREAT2:
*--#] TREAT2:

*--#[ TREATMAIN:
*--#] TREATMAIN:

*--#[ d3l79:
        ((1)
        *Dg(nu3,nu4,p5)
        *Dg(nu5,nu6,-p6)
        *Dg(nu1,nu2,-q1,-p1)
        *S(nu2,-p1m,nu5,-p4m,nu4,-p3m,nu6,-p2m,nu3,-p1m,nu1)
        *1);

        #define TOPOLOGY "O4"
*--#] d3l79:
\end{verbatim}}

\noindent
Note that in the fold {\tt TREAT0} the vector part gets multiplied by {\tt a}
in order to distinguish $\Sigma_V$ from $\Sigma_S$ in the final result.
The input for the diagram corresponds to the one pictured in
Fig.~\ref{fig:fp}.  
The complete output appearing on the screen and the result can be found in the
Appendix~\ref{app:testrun}.

\begin{figure}[ht]
  \begin{center}
  \leavevmode
      \epsfxsize=5cm
      \epsffile[177 292 435 448]{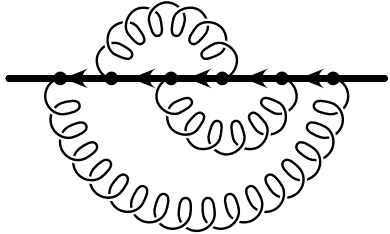}\\
      \caption[]{\label{fig:fp}Sample diagram contributing to the
  fermion propagator. Solid and looped lines represent quarks
  and gluons, respectively.
        }
  \end{center}
\end{figure}


\section*{Acknowledgments}

It is a pleasure for me to thank
K.G.~Chetyrkin, R.~Harlander, J.H.~K\"uhn and T.~Seidensticker
for many discussions and useful advice within the recent years
when {\tt MATAD} was developed.
This work was supported in part by DFG under Contract Ku 502/8-1
({\it DFG-Forschergruppe ``Quantenfeldtheorie, Computeralgebra und
  Monte-Carlo-Simulationen''})
and by SUN Microsystems through Academic Equipment Grant No.~14WU0148.
I am grateful to the department of Theoretical Particle Physics
of the University of Karls\-ruhe for the pleasant atmosphere 
during a visit when a major part of this project was carried out.


\section*{Appendix}

\begin{appendix}

\section{\label{app:ibp}Integration-by-parts}

The method of integration-by-parts plays a fundamental role in the computation
of multi-loop diagrams~\cite{CheTka81}.
For this reason we want to demonstrate its underlying
idea by considering a typical example.

The integration-by-parts algorithm uses the
fact that the $D$-dimensional integral over a total derivative is equal
to zero:
\begin{eqnarray}
  \int{\rm d}^D p {\partial\over \partial p^\mu} f(p,\ldots) &=& 0\,.
  \label{eqipgen}
\end{eqnarray}
By explicitly performing the differentiations one obtains recurrence
relations connecting Feynman integrals of different complexity.
The proper combination of different recurrence relations allows
any Feynman integral (at least single-scale ones) to be reduced to a
small set of so-called master integrals. The latter ones have to be
evaluated only once and for all, either analytically or numerically.

\begin{figure}[ht]
  \begin{center}
  \leavevmode
      \epsfxsize=5cm
      \epsffile[189 293 423 500]{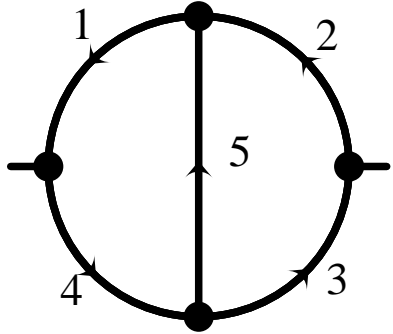}\\
      \caption[]{\label{fig:triangle}\sloppy
        Two-loop master diagram. The arrows denote the direction of
        momentum flow.
        }
  \end{center}
\end{figure}

Let us for definiteness
consider the scalar two-loop diagram of Fig.~\ref{fig:triangle}.
The corresponding Feynman integral shall be denoted by
\begin{equation}
I(n_1,\ldots,n_5) = \int{{\rm d}^D p\over (2\pi)^D}{{\rm d}^D k\over (2\pi)^D}
{1\over (p_1^2 + m_1^2)^{n_1}\cdots(p_5^2+m_5^2)^{n_5}}\,,
\end{equation}
where $p_1,\ldots,p_5$ are combinations of the loop momenta $p,k$ and
the external momentum $q$ (we work in Euclidean space here).
$n_1,\ldots,n_5$ are called the indices of the integral.  Consider the
sub-loop defined by the
lines 2, 3 and 5, and take its loop momentum to be
$p=p_5$. If we then apply the operator
$(\partial/\partial p_5)\cdot p_5$ to the {\it integrand} of $I$, we obtain a
relation of the 
form (\ref{eqipgen}), where
\begin{equation}
f(p_5,\ldots) = \frac{p_5^{\mu}}
  {(p_5^2+m_5^2)^{n_5} (p_2^2+m_2^2)^{n_2} (p_3^2+m_3^2)^{n_3}}
\,.
\end{equation}
Performing the differentiation and using momentum conservation at each
vertex one derives the following equation:
\begin{eqnarray}
\Big[ 
    - n_3 {\bf 3^+}\left({\bf 5^-}-{\bf 4^-}+m_4^2-m_5^2-m_3^2\right)
    - n_2 {\bf 2^+}\left({\bf 5^-}-{\bf 1^-}+m_1^2-m_5^2-m_2^2\right)
\nonumber\\\mbox{}
    + D-2n_5-n_3-n_2+2n_5 m_5^2 {\bf 5^+} 
\Big]\, I(n_1,\ldots,n_5) \,\,=\,\, 0
\,,
\label{eqtrianglerec}
\end{eqnarray}
where the operators ${\bf 1^{\pm}}, {\bf 2^{\pm}}, \ldots$ are used in
order to raise and lower the indices: ${\bf I^{\pm}}I(\ldots,n_i,\ldots)
= I(\ldots,n_i\pm 1,\ldots)$.  In Eq.~(\ref{eqtrianglerec}), generally
referred to as the triangle rule, it is understood that the operators to
the left of $I(n_1,\ldots,n_5)$ are applied {\em before} integration.  If the
condition $m_5=0, m_3=m_4$ and $m_1=m_2$ holds, increasing one index
always means to reduce another one.  Therefore this recurrence relation
may be used to shift the indices $n_1$, $n_4$ or $n_5$ to zero which leads
to much simpler integrals.

\begin{figure}[th]
\leavevmode
 \begin{center}
 \begin{tabular}{cccccc}
   \epsfxsize=2.5cm
   \parbox{1cm}{\epsffile[189 293 423 499]{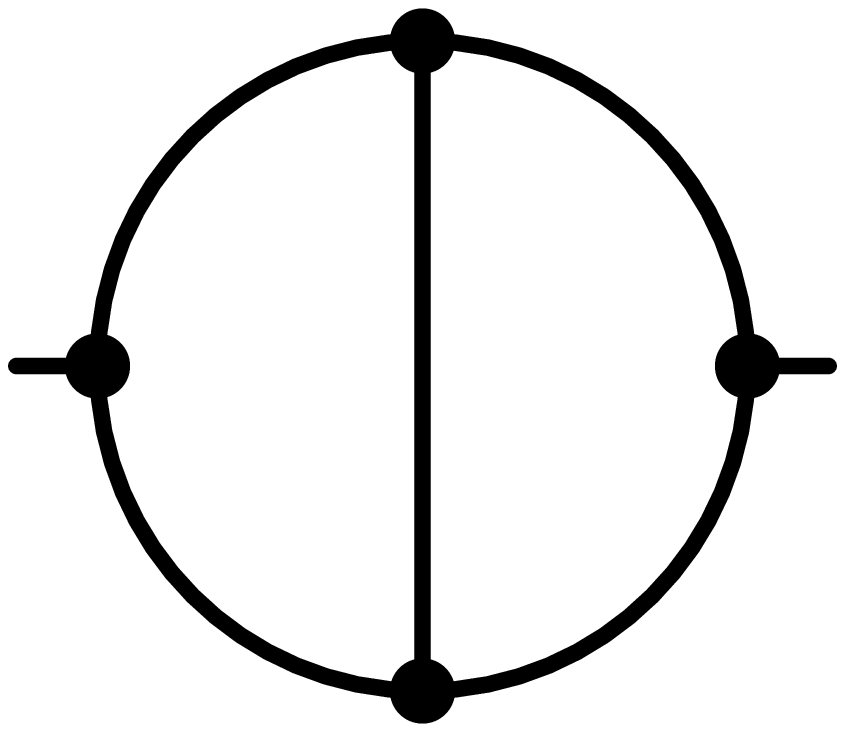}}
&
$\displaystyle =\frac{1}{\varepsilon}\Bigg[$
&
   \epsfxsize=2.5cm
   \parbox{1cm}{\epsffile[189 293 423 499]{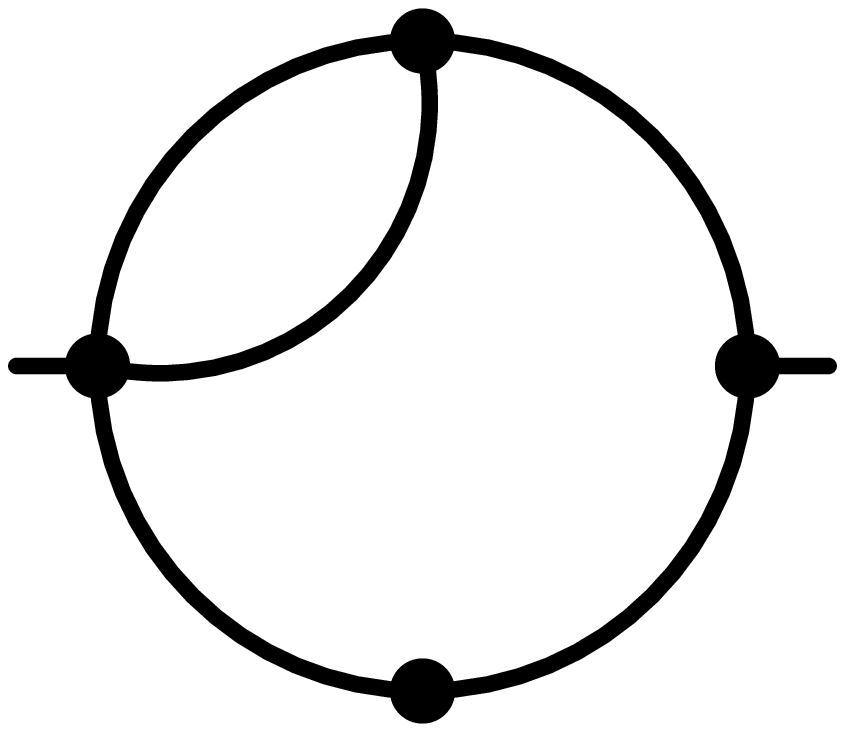}}
&
---
&
   \epsfxsize=2.5cm
   \parbox{1cm}{\epsffile[189 293 423 499]{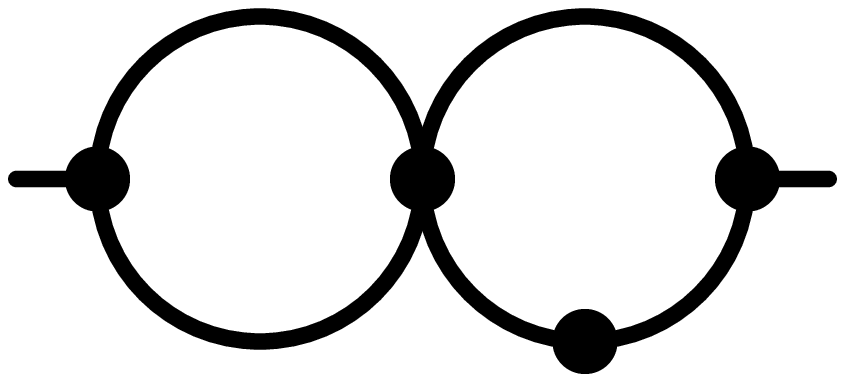}}
&
$\displaystyle \Bigg]$
 \end{tabular}
 \caption[]{\label{fig:2loopIP}\sloppy
          Symbolic equation resulting from Eq.~(\ref{eqrecI})
          applied to the diagram $I(1,1,1,1,1)$. The dot indicates that
          the respective denominator appears twice.
         }
 \end{center}
\end{figure}

The triangle rule constitutes an important building block for the
general recurrence relations. The strategy is to combine several
independent equations of the kind~(\ref{eqtrianglerec}) in order to
arrive at relations connecting one complicated integral to a set of
simpler ones. For example, while the direct evaluation of even the
completely massless case for the diagram in Fig.~\ref{fig:triangle} is
non-trivial, application of the triangle rule (\ref{eqtrianglerec})
leads to
\begin{equation}
I(n_1,\ldots,n_5) \,=\, \frac{1}{D-2n_5-n_2-n_3}
\Big[
n_2{\bf 2^+}\left({\bf 5^-} - {\bf 1^-}\right)
+
n_3{\bf 3^+}\left({\bf 5^-} - {\bf 4^-}\right)
\Big]\,I(n_1,\ldots,n_5)
\,.
\label{eqrecI}
\end{equation}
Repeated application of this equation reduces one of the indices $n_1$,
$n_4$ or $n_5$ to zero. For example, for the simplest case
($n_1=n_2=\ldots=n_5=1$) one obtains the equation pictured in
Fig.~\ref{fig:2loopIP}: The non-trivial diagram on the l.h.s.\ is
expressed as a sum of two quite simple integrals which can be solved by
applying the one-loop formula. This
example also shows a possible trap of the integration-by-parts
technique. In general its application introduces artificial
$1/\varepsilon$ poles which cancel only after combining all terms. They
require the expansion of the individual terms up to sufficiently high
powers in $\varepsilon$ in order to obtain, for example, the finite part
of the original diagram.  This point must carefully be respected in
computer realizations of the integration-by-parts algorithm: One must
not cut the series at too low powers because then the result goes wrong;
keeping too many terms, on the other hand, may intolerably slow down the
performance.

In our example, the l.h.s.\ in Fig.~\ref{fig:2loopIP} is finite, each term
on the r.h.s., however, develops $1/\varepsilon^2$ poles. The first
three orders in the expansion for $\varepsilon\to 0$ cancel, and the
${\cal O}(\varepsilon)$ term of the square bracket, together with the
$1/\varepsilon$ in front of it, leads to the well-known result
(omitting factors $1/16\pi^2$):
$I(1,1,1,1,1)=6\zeta(3)/q^2$, where $q$ is the external momentum.

In general, the successive application of recurrence relations generates
a huge number of terms out of a single diagram.  Therefore, a calculation
carried out by hand becomes very tedious and the use of computer algebra
is essential.


\section{\label{app:topfile}The topology files}

This part of the appendix provides a complete list of those massive/massless
combinations which are implemented in the topology files
{\tt inc/TOPOLOGY/topXY}.
In the following tables for all three-loop topologies of Fig.~\ref{fig:intop}
the lines are listed which have to be massive.
All other lines may be massless or absent.
In some cases some of the lines have to be completely absent, i.e. it is even
not allowed for them to be massless. This is explicitly
specified in the columns ``absent''.

\begin{center}

\begin{tabular}{| l|l || l|l || l|l || l|l|}
\hline
\multicolumn{8}{|l|}{\verb|L1|}\\ 
\hline
massive\hspace{.45cm} & absent & massive\hspace{.45cm} & absent & 
massive\hspace{.45cm} & absent & massive\hspace{.45cm} & absent \\
\hline
1 & --- & 2 & --- & 1,2 & --- & &\\
\hline
\end{tabular}
\begin{tabular}{| l|l || l|l || l|l || l|l|}
\hline
\multicolumn{8}{|l|}{\verb|T1|}\\ 
\hline
massive\hspace{.45cm} & absent & massive\hspace{.45cm} & absent & 
massive\hspace{.45cm} & absent & massive\hspace{.45cm} & absent \\
\hline
1,2 & --- & 1,2,4 & --- & 1,2,3,4 & --- & 1,4 & ---\\
2,4 & --- & 1 & --- & 2 & --- & 1,2,5 & ---\\
\hline
\end{tabular}
\begin{tabular}{| l|l || l|l || l|l || l|l|}
\hline
\multicolumn{8}{|l|}{\verb|T2|}\\ 
\hline
massive\hspace{.45cm} & absent & massive\hspace{.45cm} & absent & 
massive\hspace{.45cm} & absent & massive\hspace{.45cm} & absent \\
\hline
1,3 & --- & 2 & --- & 1,2,3 & --- & 2,4 & ---\\
3,4 & --- & 3 & --- & 1,2 & --- & &\\
\hline
\end{tabular}
\begin{tabular}{| l|l || l|l || l|l || l|l|}
\hline
\multicolumn{8}{|l|}{\verb|BE|}\\ 
\hline
massive\hspace{.45cm} & absent & massive\hspace{.45cm} & absent & 
massive\hspace{.45cm} & absent & massive\hspace{.45cm} & absent \\
\hline
1,2,3,4,5 & --- & 1,2,3 & --- & 4,5 & --- & &\\
\hline
\end{tabular}
\begin{tabular}{| l|l || l|l || l|l || l|l|}
\hline
\multicolumn{8}{|l|}{\verb|BU|}\\ 
\hline
massive\hspace{.45cm} & absent & massive\hspace{.45cm} & absent & 
massive\hspace{.45cm} & absent & massive\hspace{.45cm} & absent \\
\hline
1,3,6,7 & --- & 4,5,6,7 & --- &  & & &\\
\hline
\end{tabular}
\begin{tabular}{| l|l || l|l || l|l || l|l|}
\hline
\multicolumn{8}{|l|}{\verb|LA|}\\ 
\hline
massive\hspace{.45cm} & absent & massive\hspace{.45cm} & absent & 
massive\hspace{.45cm} & absent & massive\hspace{.45cm} & absent \\
\hline
1,2,3,4,5,6 & --- & 1,3,4,6,7,8 & --- & 1,6,7 & --- & 3,4,8 & ---\\
1,2,3 & --- & 1,3,7,8 & 4,5,6 & 1,7 & 4,5,6 & 3,8 & 4,5,6\\
\hline
\end{tabular}
\begin{tabular}{| l|l || l|l || l|l || l|l|}
\hline
\multicolumn{8}{|l|}{\verb|NO|}\\ 
\hline
massive\hspace{.45cm} & absent & massive\hspace{.45cm} & absent & 
massive\hspace{.45cm} & absent & massive\hspace{.45cm} & absent \\
\hline
1,2,3,4,5,6 & --- & 1,2,3 & --- &  & & &\\
\hline
\end{tabular}
\begin{tabular}{| l|l || l|l || l|l || l|l|}
\hline
\multicolumn{8}{|l|}{\verb|N2|}\\ 
\hline
massive\hspace{.45cm} & absent & massive\hspace{.45cm} & absent & 
massive\hspace{.45cm} & absent & massive\hspace{.45cm} & absent \\
\hline
1,2,3,4,5,6 & --- & 1,2,3,4 & --- & 5,6 & --- & 2,3 & ---\\
\hline
\end{tabular}
\begin{tabular}{| l|l || l|l || l|l || l|l|}
\hline
\multicolumn{8}{|l|}{\verb|N3|}\\ 
\hline
massive\hspace{.45cm} & absent & massive\hspace{.45cm} & absent & 
massive\hspace{.45cm} & absent & massive\hspace{.45cm} & absent \\
\hline
1,2,3,4,5 & --- & 2 & --- & 1,3,4,5 & --- & 3,4 & ---\\
3,4,5 & --- & 4,5 & --- & 1,3 & --- & 1,5 & ---\\
\hline
\end{tabular}
\begin{tabular}{| l|l || l|l || l|l || l|l|}
\hline
\multicolumn{8}{|l|}{\verb|O1|}\\ 
\hline
massive\hspace{.45cm} & absent & massive\hspace{.45cm} & absent & 
massive\hspace{.45cm} & absent & massive\hspace{.45cm} & absent \\
\hline
1,2,3,4,6,7 & --- & 1,2,3,4 & --- & 6,7 & --- & 1,2,6,7 & ---\\
1,2 & --- & 1,5,6 & 3,4 & 1,5,7 & 3,4 & 5,6 & ---\\
5 & --- & 5,7 & --- & 1,5,7 & --- & 7 & ---\\
1,5 & --- & 1 & --- & 1,7 & --- & 6 & ---\\
1,6,7 & --- & 1,2,7 & 3,4 &  & & &\\
\hline
\end{tabular}
\begin{tabular}{| l|l || l|l || l|l || l|l|}
\hline
\multicolumn{8}{|l|}{\verb|O2|}\\ 
\hline
massive\hspace{.45cm} & absent & massive\hspace{.45cm} & absent & 
massive\hspace{.45cm} & absent & massive\hspace{.45cm} & absent \\
\hline
1,2,3,4,7 & --- & 3,4,7 & --- & 1,2 & --- & 7 & ---\\
\hline
\end{tabular}
\begin{tabular}{| l|l || l|l || l|l || l|l|}
\hline
\multicolumn{8}{|l|}{\verb|O4|}\\ 
\hline
massive\hspace{.45cm} & absent & massive\hspace{.45cm} & absent & 
massive\hspace{.45cm} & absent & massive\hspace{.45cm} & absent \\
\hline
1,2,6,7 & --- & 1,2,6 & --- & 7 & --- & 1,3,5,6 & 7\\
1,3,5 & --- & 3,5 & --- & 1,5 & --- & 5 & ---\\
1,3,5,7 & --- & 3,5,7 & --- & 1,2 & --- & 1 & ---\\
2 & --- & 1,6 & --- & 2,6 & --- & 1,3,7 & 6\\
3,7 & 6 & 2,6 & --- & 1,3,5,7 & 6 & 1,5,7 & 6\\
1,2,7 & 6 & 2,3 & 7 & 1,2,3 & 7 & 3,4 & 7\\
3,4,5 & 7 & 1,3,4 & 7 & 1,2,5 & 7 & 1,3,4,5,6 & 7\\
1,4,5,6 & 7 & 1,3,4,5 & 7 & 1,4,5 & 7 & 1,4 & 7\\
3,4,6 & 7 & 1,2,3,4 & 7 & 2,3,4 & 7 & 1,2,3,4,5 & 7\\
1,2,4,5 & 7 & 1,3,4,6 & 7 & 1,2,3,5 & 7 & 2,3,5 & 7\\
3,4,5,6 & 7 & 4,5,6 & 7 & 2,3,4,5 & 7 & 2,4,5 & 7\\
2,4 & 7 & 1,2,4 & 7 & 1,5 & 7 & 1,3 & 7\\
3 & 7 & 2,5 & 7 & 1,2 & 7 & 4,5 & 7\\
4 & 7 & 1,2,4,5,6 & 7 & 2,4,5,6 & 7 & 6 & 7\\
4,6 & 7 & 3,5,6 & 7 & 3,6 & 7 & 1,5,6 & 7\\
2,5,6 & 7 & 1,3,6 & 7 & 1,2,5,6 & 7 & 1,2,4,6 & 7\\
2,4,6 & 7 & 5,6 & 7 & 1,4,6 & 7 & &\\
\hline
\end{tabular}
\begin{tabular}{| l|l || l|l || l|l || l|l|}
\hline
\multicolumn{8}{|l|}{\verb|Y1|}\\ 
\hline
massive\hspace{.45cm} & absent & massive\hspace{.45cm} & absent & 
massive\hspace{.45cm} & absent & massive\hspace{.45cm} & absent \\
\hline
1,2,3,6 & --- & 1,3 & --- & 3 & --- & 3,5,6 & ---\\
\hline
\end{tabular}
\begin{tabular}{| l|l || l|l || l|l || l|l|}
\hline
\multicolumn{8}{|l|}{\verb|Y2|}\\ 
\hline
massive\hspace{.45cm} & absent & massive\hspace{.45cm} & absent & 
massive\hspace{.45cm} & absent & massive\hspace{.45cm} & absent \\
\hline
1,2,3,5 & --- & 2 & --- & 1,3,5 & --- & 3 & ---\\
5 & --- & 1,3 & --- & 1,5 & --- & &\\
\hline
\end{tabular}
\begin{tabular}{| l|l || l|l || l|l || l|l|}
\hline
\multicolumn{8}{|l|}{\verb|Y3|}\\ 
\hline
massive\hspace{.45cm} & absent & massive\hspace{.45cm} & absent & 
massive\hspace{.45cm} & absent & massive\hspace{.45cm} & absent \\
\hline
1,2,3,5 & --- & 1,3,5 & --- & 2 & --- & 5 & ---\\
3,5 & --- & 1,3 & --- &  & & &\\
\hline
\end{tabular}

\end{center}


\section{\label{app:not}Some details on the notation}

\subsection{\label{app:notin}Notation of the input}

In this subsection we describe the notation to be used for the input.

On one side of the following expressions
the {\tt FORM} notation is used
whereas the other side displays the corresponding mathematical terms.
It is always assumed that $p_i$ ($i=1,\ldots,9$)
is a loop momentum and $q_i$ ($i=1,\ldots,3$) is a small
momentum in which an expansion is performed.

If scalar integrals are to be computed with {\tt MATAD}
the following notation has to be used:
\begin{eqnarray}
  \begin{array}{llll}
    \mbox{massless lines:} & \frac{1}{p_1^2},\frac{1}{p_2^2},\ldots &
    \longrightarrow & \mbox{\verb|1/p1.p1|, \verb|1/p2.p2|},\ldots
    \\
    & \frac{1}{-(p_1+q_1)^2},\ldots &
    \longrightarrow & \mbox{\verb|Dl(p1,q1)|},\ldots
    \\
    \mbox{massive lines:} & \frac{1}{M^2-p_1^2},\frac{1}{M^2-p_2^2},\ldots &
    \longrightarrow & \mbox{\verb|s1m|, \verb|s2m|},\ldots  
    \\
    & \frac{1}{M^2-(p_1+q_1)^2},\ldots &
    \longrightarrow & \mbox{\verb|Dh(p1,q1)|},\ldots    
\end{array}
\end{eqnarray}
Fermions are treated with the help of the (non-commutative)
function \verb|S| and the function \verb|Dg| can be used for the
gluon propagator:
\renewcommand\arraystretch{2}
\begin{eqnarray}
  \begin{array}{llllll}
    \verb|S(mu)|
    &=& 
    \gamma^\mu
    \,,
    &
    \verb|S(p1)|
    &=& 
    \frac{p_1\hspace{-.7em}/}{-p_1^2}
    \,,
    \\
    \verb|S(p1m)|
    &=& 
    \frac{M+p_1\hspace{-.7em}/}{M^2-p_1^2}
    \,,
    &
    \verb|S(q1,p1)|
    &=& 
    \frac{p_1\hspace{-.7em}/\, + q_1\hspace{-.7em}/}{-(p_1+q_1)^2}
    \,,
    \\
    \verb|S(q1,p1m)|
    &=& 
    \frac{M+p_1\hspace{-.7em}/\, + q_1\hspace{-.7em}/}{M^2-(p_1+q_1)^2}
    \,,
    \\
    \verb|Dg(mu,nu,p1)|
    &=& 
    \frac{-g^{\mu\nu} - \xi \frac{p_1^\mu p_1^\nu}{-p_1^2}}{-p_1^2}
    \,,
    &
    \verb|Dg(mu,nu,q1,p1)|
    &=& 
    \frac{-g^{\mu\nu} - \xi \frac{(p_1+q_1)^\mu (p_1+q_1)^\nu}
      {-(p_1+q_1)^2}}{-(p_1+q_1)^2}
    \,.
  \end{array}
  \label{eq:SDg}
\end{eqnarray}
\renewcommand\arraystretch{1}

\noindent
Instead of \verb|S| also \verb|SS|, \verb|SSS| or \verb|SSSS| may be
used which is useful in case more than one fermion line is involved.
The vertices between three gluons and gluon and ghosts 
are implemented through the following functions (where the colour
factors are not taken into account):
\begin{eqnarray}
  \begin{array}{lll}
    \verb|V3g(i1,p1,i2,p2,i3,p3)|
    &=&
    \left(p_2- p_1\right)^{i_3} g^{i_1 i_2}
    +\left(p_3- p_2\right)^{i_1} g^{i_2 i_3}
    +\left(p_1- p_3\right)^{i_2} g^{i_3 i_1}
    \,,
    \\
    \verb|Vgh(i1,p1)| &=& -p_1^{i_1}
    \,.
  \end{array}
  \label{eq:V3g}
\end{eqnarray}
If small momenta are present they can simply be added to $p_1$,
$p_2$ or $p_3$.
In analogy to the expansion in a small momentum it is possible to introduce
functions where an expansion in small masses can be performed.

In the above expressions 
instead of \verb|q1| also \verb|q2| or \verb|q3| may be chosen as 
momenta in which an expansion is done.
For the loop momenta \verb|p1|, \ldots, \verb|p9| is allowed.

For the Feynman rules given above the user has to check the
consistency with own sign conventions. In particular all
unnecessary factors like, e.g., the strong coupling constant are
omitted.
Also the imaginary unit, $i$, is suppressed. Furthermore the Feynman
rules are chosen in such a way that the vertices are proportional to
$i$ and all propagators are proportional to $1/i$ which leads to
convenient cancellations.
Thus each diagram can be written as 
\begin{eqnarray}
  i\left(\frac{1}{i} \int\frac{{\rm d}^D k_j}{\left(2\pi\right)^D}
  \right)^l
  \times[\mbox{expression formed by the Feynman rules
  of~(\ref{eq:SDg}) and~(\ref{eq:V3g}) }]
  \,,
\end{eqnarray}
where $l$ is the number of loops. The factor $1/i$ in 
front of the integration momenta disappears after the Wick rotation.
The overall factor $i$, which occurs for all diagrams and which is
independent of the number of loops, is omitted.

\subsection{Notation and conventions of the output}

At this point some words about the notation of the output of {\tt MATAD}
are in order.
As already mentioned, the final expressions are
expanded in $\varepsilon$
where --- following general $\overline{\rm MS}$ conventions ---
the factors $\gamma_E$ and $\ln4\pi$ have been dropped.
Furthermore the factors $(1/16\pi^2)^l$ and $(1/M^{2\varepsilon})^l$
where $l$ is the number of loops are omitted.
Concerning the remaining symbols the translation is given in the following
table:
\begin{center}
  \begin{tabular}{ll}
    $\varepsilon$ & \verb|ep| \\
    Euclidean external momenta  & \verb|Q, Q1, Q2,|~\ldots \\
    $\zeta_2, \zeta_3, \ldots$ (Rieman zeta function) & 
    \verb|z2, z3,|~\ldots \\
    mass appearing in the integrals  & \verb|M| \\
    parametrization of the 
    finite parts
    \\
    of the master integrals & \verb|D5, D4,|~\ldots
    (see Appendix~\ref{app:master})
  \end{tabular}
\end{center}
Note that all momenta which occur in the output have to be interpreted
in Euclidean space.

\subsection{\label{app:master}Master integrals}

{\tt MATAD} does not insert the complete 
finite parts of the master integrals
automatically. Instead they are parameterized by the symbols
{\tt D5}, {\tt D4}, {\tt DM}, {\tt DN}, {\tt B4}, ...
where the notation is adopted from the one introduced in Fig.~\ref{fig:batop}.
In the following list one can find the corresponding analytical 
expressions~\cite{Bro92,Bro98,FleKal99,CheSte00,DavTau93}\footnote{Note that
there is a misprint in Eq.~(22) of Ref.~\cite{CheSte00}:
all four appearing Clausen functions must be squared and the
coefficient of $\zeta_4$ in the second to last equation must read
``$-22$'' instead of ``$-31/2$''.}

\begin{eqnarray}
  \verb|D6| &=& 
                  6\zeta_3
                -17 \zeta_4 
                -4 \zeta_2\ln^2 2
                +\frac{2}{3}\ln^4 2
                + 16 \mbox{Li}_4\left(\frac{1}{2}\right) 
                -4 \left[{\rm Cl}_2\left(\frac{\pi}{3}\right)\right]^2 
  \,,
  \nonumber\\
  \verb|D5| &=& 6\zeta_3-\frac{469}{27}\zeta_4
  +\frac{8}{3}\left[{\rm Cl}_2\left(\frac{\pi}{3}\right)\right]^2 
  -16\sum_{m>n>0} \frac{(-1)^m \cos(2\pi n/3)}{m^3n}
  \nonumber\\&\approx&
  -8.2168598175087380629133983386010858249695
  \,,
  \nonumber\\
  \verb|D4| &=& 
  6\zeta_3
  -\frac{77}{12}\zeta_4 
  -6 \left[\mbox{Cl}_2\left(\frac{\pi}{3}\right) \right]^2 
  \,,
  \nonumber\\
  \verb|D3| &=&
  6\zeta_3
  -\frac{15}{4}\zeta_4 
  -6 \left[\mbox{Cl}_2\left(\frac{\pi}{3}\right) \right]^2 
  \,,
  \nonumber\\
  \verb|DM| &=& 
  6\zeta_3
  -\frac{11}{2}\zeta_4 
  -4 \left[\mbox{Cl}_2\left(\frac{\pi}{3}\right) \right]^2 
  \,,
  \nonumber\\
  \verb|DN| &=& 
  6\zeta_3
  - 4\zeta_2\ln^2 2
  + \frac{2}{3}\ln^4 2  
  - \frac{21}{2}\zeta_4  
  + 16 \mbox{Li}_4\left(\frac{1}{2}\right)
  \,,
  \nonumber\\
  \verb|B4| &=& 
  - 4 \zeta_2 \ln^2 2
  + \frac{2}{3}\ln^4 2 
  - \frac{13}{2} \zeta_4
  + 16 \mbox{Li}_4\left(\frac{1}{2}\right) 
  \,,
  \nonumber\\
  \verb|E3| &=&
  - \frac{139}{3} 
  - \frac{\pi\sqrt{3} \ln^2 3}{8} 
  - \frac{17\pi^3\sqrt{3}}{72}
  - \frac{21}{2}\zeta_2
  + \frac{1}{3}\zeta_3
  \nonumber\\&&\mbox{}
  + 10\sqrt{3}{\rm Cl}_2\left(\frac{\pi}{3}\right)
  - 6\sqrt{3} {\rm Im}\left[ {\rm
     Li}_3\left(\frac{e^{-i\pi/6}}{\sqrt{3}}\right)\right] 
  \,,
  \nonumber\\
  \verb|S2| &=& \frac{4}{9\sqrt{3}} \mbox{Cl}_2\left(\frac{\pi}{3}\right)
  \,,
  \nonumber\\
  \verb|OepS2| &=&
  - \frac{763}{32}
  - \frac{9\pi\sqrt{3}\ln^2 3}{16}
  - \frac{35\pi^3\sqrt{3}}{48}
  + \frac{195}{16}\zeta_2
  - \frac{15}{4}\zeta_3
  + \frac{57}{16}\zeta_4
  \nonumber\\&&\mbox{}
  + \frac{45\sqrt{3}}{2} {\rm Cl}_2\left(\frac{\pi}{3}\right)
  - 27\sqrt{3} {\rm Im}\left[ {\rm
     Li}_3\left(\frac{e^{-i\pi/6}}{\sqrt{3}}\right)\right]
  \,,
  \nonumber\\
  \verb|T1ep| &=&
  - \frac{45}{2} 
  - \frac{\pi\sqrt{3} \ln^2 3}{8} 
  \nonumber\\&&\mbox{}
  - \frac{35\pi^3\sqrt{3}}{216}
  - \frac{9}{2}\zeta_2
  + \zeta_3
  + 6\sqrt{3}{\rm Cl}_2\left(\frac{\pi}{3}\right)
  - 6\sqrt{3} {\rm Im}\left[ {\rm
      Li}_3\left(\frac{e^{-i\pi/6}}{\sqrt{3}}\right)\right] 
  \label{eq:master}
  \,,
\end{eqnarray}
with $\mbox{Cl}_2(x) = \mbox{Im}[\mbox{Li}_2(e^{ix})]$.
$\mbox{Li}_2$, $\mbox{Li}_3$ and $\mbox{Li}_4$ 
are the Di-, Tri- and Quadrilogarithm, respectively.
\verb|S2| both appears in the diagram of Fig.~\ref{fig:master_add} and the
two-loop diagram of Fig.~\ref{fig:tad123}
where all three lines have the same mass. Their ${\cal O}(\varepsilon)$
parts, which can contribute to the finite part of the three-loop results,
contain \verb|OepS2| and \verb|T1ep|, respectively.
We should mention that those expressions of Eq.~(\ref{eq:master})
which coincide with an existing topology (e.g.\ {\tt D5}
$\leftrightarrow$ {\tt topD5}) indeed agree with the finite part of
the corresponding master integral.
On the other hand {\tt B4}, e.g.,  comprises the complicated parts of 
the finite part of the master integral corresponding to {\tt topBN},
however, not the complete one.
\verb|D6| is not yet implemented into {\tt MATAD} and
listed for completeness only.

\subsection{\label{app:settings}Parameters and switches in {\tt main.<prb>}}

In the following a brief description of the switches in the file
{\tt main<prb>} is given. In all cases the {\tt FORM} syntax reads
\verb|#define <VAR> "<VAL>"| where \verb|<VAR>| is the variable to be defined
and \verb|<VAL>| the corresponding value.

It is required to define at least the following five variables:
\begin{description}
\item[{\tt DIAGRAM:}]
  name of the diagram to be computed.
\item[{\tt FOLDER:}]
  the file containing the special treat files and the diagrams
  is called {\tt 'FOLDER'.dia}.
\item[{\tt GAUGE:}]
  determines the choice for the gauge parameter.
  The variable $\xi$ as defined in the function
  \verb|Dg| in Eq.~(\ref{eq:SDg}) is replaced by the value of 
  {\tt GAUGE}.
  In particular \verb|#define GAUGE "0"| corresponds to Feynman gauge and
  with \verb|#define GAUGE "xi"| the calculation is performed for general
  gauge parameter.
\item[{\tt POWER:}]
  determines the depth of the expansion in the small quantities.
\item[{\tt PRB:}]
  name of problem. {\tt PRB} corresponds to the name of the fold
  where the problem-dependent files can be found.
\end{description}
The definition of the remaining variables is optional:
\begin{description}
\item[{\tt BNRECOLD:}]
  if this variable is defined the original recurrence procedure
  of Ref.~\cite{Bro92} is used for the topology {\tt BN}.
\item[{\tt CUT:}]
  determines the depth of the expansion in $\varepsilon$ of the final result.
  At one-, two- and three-loop order at most the terms 
  of order $\varepsilon^2$, $\varepsilon^1$, respectively, $\varepsilon^0$ 
  are reliable.
  The default value is ``2''.
\item[{\tt DALA12:}]
  expansion in $q_1 q_2$. If this variable is set
  $q_1^2$ and $q_2^2$ are set to zero and powers in $q_1 q_2$
  are factored out.
  Currently only the expansion up to order 
  $(q_1 q_2)^4$ is implemented.
  Note that only positive powers of $q_1 q_2$ can be treated.
\item[{\tt DALAQN:}]
  apply d'Alembert operator,
  $\Box_q=\partial/\partial q^\mu \partial/\partial q_\mu$,
  in order to factor out powers 
  in $q^2$. In the argument of {\tt DALAQN} the momentum is specified with
  respect to which the derivatives are performed.  
\item[{\tt NOR:}]
  is an abbreviation for Number Of Recursions.
  As the naive use of the {\tt repeat}--{\tt endrepeat} construction 
  significantly slows down the performance the most complicated procedures are
  buffered by a \verb|#do|--\verb|#enddo| construction. 
  {\tt 'NOR'} constitutes the upper bound of the do-loop.
  The default value is ``10''.
\item[{\tt PROBLEM0/1/2/MAIN:}]
  If one of these variables is set a special treat file is read at the 
  corresponding position. The value has to agree with the one defined in
  the file  \verb|<prb>.dia|.
\item[{\tt TIME:}]
  If this variable is set the statistics is printed at various steps 
  of the calculation.
\end{description}

There are more switches in {\tt inc/main.gen}. However, they should not be
modified as most of them are in an experimental stage and not sufficiently
tested.


\section{\label{app:files}List of files}

In this appendix we provide a list of all files belonging to {\tt MATAD}.
The directory {\tt inc} contains essentially the include-files.
In particular some of the procedures are collected in files
which are included at the very beginning.
The tables for the topology {\tt BN} are located in {\tt
inc/TABLEDAL} and {\tt inc/TABLEREC}. The directories {\tt inc/TREAT} and 
{\tt inc/TOPOLOGY} essentially contain the files treating the
individual (input and basic) topologies
and in the folder {\tt prc} some auxiliary procedures are collected.

{\footnotesize
\begin{verbatim}
form.set  inc/  matadform  prc/  problems/

inc:
TABLEDAL/  bnm2m          expandDr    nomBM          reduceBM_2  tblBN
TABLEREC/  declare.matad  expepgam    nomBN          reduceBN
TOPOLOGY/  denoexp        expnomdeno  nomdecomBN     reduceBN1
TREAT/     expandBN       main.gen    recursion_2    reduceBN_2
bnbm2prc   expandBNM      matad.info  redcut         symmetryBM
bnbmprc    expandBNM_2    matminprc   redcutnomdeno  symmetryBN

inc/TABLEDAL:
BNd.tbl  BNd0.tbl

inc/TABLEREC:
BN.tbl  BNn1.tbl  BNn1n2.tbl  BNn2.tbl

inc/TOPOLOGY:
topBE    topBN    topBU  topE3  topM2  topN3  topO4.add  topY3
topBM    topBN1   topD4  topE4  topM3  topNO  topT1      topempty
topBM1   topBN2   topD5  topL1  topM4  topO1  topT2
topBM2   topBN3   topDM  topLA  topM5  topO2  topY1
topBM_2  topBN_2  topDN  topM1  topN2  topO4  topY2

inc/TREAT:
treat.dala12  treatbm2   treatbn2   treatd5  treate4  treatm4
treat.dalaav  treatbm_2  treatbn3   treatdm  treatm1  treatm5
treatbm       treatbn    treatbn_2  treatdn  treatm2  treatn1
treatbm1      treatbn1   treatd4    treate3  treatm3  treatt1

prc:
aver.prc   cutep.prc     dalaqn.prc  difvecsc.prc  solveS.prc  treat.prc
aver1.prc  dala12sc.prc  dalasc.prc  pochtabl.prc  tabBN.prc
\end{verbatim}}

\subsection*{Files taken over from {\tt MINCER}}

{\footnotesize
\begin{verbatim}
one.prc       simplify.prc  finish.prc     tabtwo.prc
accu.prc      dotwo.prc     newtwo.prc     two.prc
triangl2.prc  triangle.prc  pochtabl.prc
\end{verbatim}}


\section{\label{app:testrun}Fermion propagator: output}

In this appendix we present the complete output of the example discussed in
Section~\ref{sub:ferm}.
Calling {\tt MATAD}

{\footnotesize
\begin{verbatim}
> matadform problems/fp/mainfp
\end{verbatim}}

\noindent
leads to

{\footnotesize
\begin{verbatim}
    FORM version 2.3 Apr 24 1997

    *
    * mainfp
    *
    #define PRB "fp"
    #define PROBLEM0 "1"
    #define DALAQN "q1"
    #define GAUGE "0"
    #define POWER "1"
    #define CUT "0"
    #define FOLDER "fp"
    #define DIAGRAM "d3l79"
    #-
*~~  MATAD -- computation of MAssive TADpoles
*~~  read generic main file
*~~  read diagram
    G dia=
    #include problems/'PRB'/'FOLDER'.dia # 'DIAGRAM'
            ((1)
            *Dg(nu3,nu4,p5)
            *Dg(nu5,nu6,-p6)
            *Dg(nu1,nu2,-q1,-p1)
            *S(nu2,-p1m,nu5,-p4m,nu4,-p3m,nu6,-p2m,nu3,-p1m,nu1)
            *1);
    
            #define TOPOLOGY "O4"
    *--#] d3l79:
    #-
*~~  Treat the traces
*~~  Include special treat-file 0
*~~  Feynman rules for vertices and propagators:
*~~    gluon-ghost-ghost-vertex
*~~    3-gluon-vertex
*~~    gluon propagator
*~~    ghost propagator
*~~  expand denominators
*~~  Dh
*~~  Dl
*~~  1
*~~  1
*~~  Change notation to p1,p2,...
*~~  Trace 1
*~~  Trace 2
*~~  Trace 3
*~~  Trace 4
*~~  treat DL(x)
*~~  Do Wick-rotation
*~~  Apply d Alembertian w.r.t. q1
*~~  average done
*~~  Expand Dr(p,q)
*~~  1
*~~  q_i -> Q_i
*~~  include TOPOLOGY-file
*~~  this is topO4
*~~  Recursion of type d5
*~~  this is topD5
*~~  numerator
*~~  do recursion
*~~  - done
*~~  Recursion of type d4
*~~  this is topD4
*~~  numerator
*~~  do recursion
*~~  - done
*~~  Recursion of type dm
*~~  this is topDM
*~~  numerator
*~~  do recursion
*~~  - done
*~~  Recursion of type dn
*~~  this is topDN
*~~  numerator
*~~  do recursion
*~~  - done
*~~  Recursion of type e4
*~~  this is topE4
*~~  numerator
*~~  do recursion
*~~  - done
*~~  Recursion of type e3
*~~  this is topE3
*~~  numerator
*~~  do recursion
*~~  - done
*~~  Recursion of type bn_2
*~~  this is topBN_2
*~~  numerator
*~~  do recursion
*~~  Use table for BN
*~~  - done
*~~  Recursion of type bn1
*~~  this is topBN1
*~~  numerator
*~~  do recursion
*~~  - done
*~~  Recursion of type bn2
*~~  this is topBN2
*~~  numerator
*~~  do recursion
*~~  - done
*~~  Recursion of type bn3
*~~  this is topBN2
*~~  numerator
*~~  do recursion
*~~  - done
*~~  Recursion of type bm_2
*~~  this is topBM_2
*~~  numerator
*~~  do recursion
*~~  - done
*~~  Recursion of type bm1
*~~  this is topBM1
*~~  numerator
*~~  do recursion
*~~  - done
*~~  Recursion of type bm2
*~~  this is topBM2
*~~  numerator
*~~  do recursion
*~~  - done
*~~  Integration of the simple integrals
*~~  Recursion of type m1
*~~  this is topM1
*~~  Recursion of type m2
*~~  this is topM2
*~~  Recursion of type m3
*~~  this is topM3
*~~  Recursion of type m4
*~~  this is topM4
*~~  Recursion of type m5
*~~  this is topM5
*~~  perform integration
*~~  Recursion of type t1
*~~  this is treatn1
*~~  Recursion of type n1
*~~  this is treatn1
*~~  Simplify
*~~  Do the "rest"-integration

Time =       5.33 sec    Generated terms =         23
            d3l79        Terms in output =         23
                         Bytes used      =        410

   d3l79 =
       + ep^-3 * (  - 8/3 - 1/3*a )

       + ep^-2 * ( 56/3 - 20/3*a )

       + ep^-1 * ( 112/3 - 16*z3 + 19/2*z2*a - 20*z2 - 97/12*a )

       + 334/3 + 1215/2*S2*a - 1620*S2 + 16*D3*a - 40*D3 - 1141/3*z3*a + 2368/
         3*z3 + 144*z4*a - 288*z4 + 57*z2*a - 156*z2 - 32*a*B4 - 77/6*a + 64*
         B4;

    save problems/'PRB'/results/'DIAGRAM'.res 'DIAGRAM';
    .end
\end{verbatim}}

\end{appendix}


\end{document}